\documentclass[prb,twocolumn,aps,showpacs]{revtex4}
\usepackage{graphicx,epsfig}% complex graphics
\usepackage{bm}% bold math\textbf{}
\usepackage{amssymb} %math symbols

\begin{document}

\title{Polaron spin echo envelope modulations in an
organic semiconducting polymer}

\author{V. V. Mkhitaryan and V. V. Dobrovitski}

\affiliation{Ames Laboratory, Iowa State University, Ames, Iowa
50011, USA}

\begin{abstract}

Theoretical treatment of the electron spin echo envelope
modulation (ESEEM) spectra from polarons in a semiconducting
$\pi$- conjugated polymer is presented. The contact hyperfine
coupling and the dipolar interaction between the polaron and
proton spins are found to have distinct contributions in the ESEEM
spectra. However, since the two contributions are spaced very
closely, and the dipolar contribution is dominant, the detection
of the contact hyperfine interaction is difficult. To resolve this
problem, a recipe of probing the contact hyperfine and dipolar
interactions {\it selectively} is proposed, and a method for
detecting the polaron contact hyperfine interaction is formulated.
The ESEEM decay due to the polaron random hopping is analyzed, and
the robustness of the method against this decay is verified.
Moreover, this decay is linked to the transport properties of
polarons, providing an auxiliary probe for the polaron transport.

\end{abstract}

\maketitle

\section{Introduction}

Over the past decades, semiconducting organic $\pi$- conjugated
small molecule and polymer materials have been become widely used
in optoelectronic devices such as light-emitting diodes and solar
cells. \cite{Forrest04, SpecIss07}
%These materials combine the useful optoelectronic properties with
%the inexpensive and relatively simple processing.
This triggered an increasing interest in the area of organic
electronics, uncovering a variety of new concepts. Remarkably, it
was established that the charge carrier spin is fundamental to
electrical and optical properties of organic semiconductors.
However, because of the extremely complex nature,
%inherently heterogeneous structure
many important aspects of the spin dynamics and underlying
microscopic mechanisms are not yet well understood. This includes
the microscopic structure of charge-carrier polaron states, and
the resulting hyperfine coupling of polaron spin to the local
magnetic environment, which is a key for understanding the
spin-dependent processes in organic semiconductors.

Optically and electrically detected magnetic resonance (ODMR and
EDMR, respectively) are highly efficient spectroscopic tools for
the investigation of microscopic properties of organic
semiconductors. \cite{Shinar} While the conventional electron spin
resonance (ESR) techniques measure the spin polarization, ODMR and
EDMR probe optically and electrically active paramagnetic states,
\cite{Cavenett81, Street82, Depinna82} which are crucial to many
organic semiconductor applications. Moreover, as the spin
polarization in organic semiconductors is typically low, ODMR and
EDMR are much more sensitive than the conventional ESR.
\cite{Stutzmann00, Lifshitz04, McCamey06}

Substantial progress in this direction was made by the pulsed EDMR
(pEDMR) experiments. \cite{Dane08, DanePRL10, DanePRB10,
Behrendes10, BoehmePRL12, Malissa, Kipp15} Unlike the continuous
wave measurements, these experiments are capable of probing the
coherent spin dynamics, and thus provide a closer view on the
spin-dependent processes. Importantly, pEDMR (and pODMR) offer the
implementation of various spin-echo based spectroscopic techniques
in the study of organic semiconductors. \cite{BoehmePRL12,
Malissa} This motivates the present theoretical study of a
spectroscopic method based on the two-pulse (Hahn) echo and
three-pulse echo sequences. \cite{Slichter}

%Important characteristics of organic semiconductors is the very
%small spin-orbit coupling, so that the polaron spin dynamics is
%governed mainly by its hyperfine interaction with the surrounding
%protons. \cite{Dediu09, ThoNatMat}

In many organic semiconductors the spin-orbital coupling is very
week, and the polaron spin dynamics is governed mainly by the
hyperfine interaction of polaron spin with the surrounding proton
spins. \cite{Dediu09, ThoNatMat} Therefore, probing the polaron
hyperfine interaction is very important. A magnetic resonance
technique widely used to investigate the hyperfine interaction of
paramagnetic centers is electron spin echo envelope modulation
(ESEEM) spectroscopy. \cite{DT, SchJ} The pEDMR implementation of
this technique, applied to organic polymer
poly[2-methoxy-5-(2$'$-ethyl-hexyloxy)- 1,4-phenylene vinylene]
(MEH-PPV), was recently reported by Malissa et al. \cite{Malissa}
Employing a version of ESEEM, the authors of Ref.
\onlinecite{Malissa} have been able to resolve the proton spectral
line in MEH-PPV and that of the deuteron and proton in partially
deuterated MEH-PPV.

In this paper we develop a theory of ESE modulations based on the
two-pulse, primary echo, and three-pulse, stimulated echo (primary
and stimulated ESEEM, respectively), in application to MEH-PPV. We
argue that the spectral lines observed in Ref.
\onlinecite{Malissa} are due to the magnetic dipolar coupling of
polaron spin with the distant protons, while the protons coupled
to the polaron spin with the contact hyperfine interaction (HFI)
are not detected. Our theory offers a way to address the distant
and the contact hyperfine protons selectively, by a proper choice
of the stimulated ESEEM parameters. Thus we propose a method of
probing the contact HFI. This also gives a valuable information on
the polaron orbital state, stipulating the contact HFI.

The paper is organized as follows. In the next Section we discuss
the hyperfine interaction between the polaron and proton spins,
particularly in polymer poly[p-phenylene vinylene] (PPV) and its
derivative, MEH-PPV.
%and its semiclassical description in terms of random static
%magnetic fields.
The derivation of ESEEM is given in Section \ref{SecDeriv}. In
Section \ref{SecOrDis} we analyze the effect of random
orientations of polymer chains. The polaron hopping and the
resulting ESE modulation decay is considered in
Section~\ref{SecHopping}. We discuss our results in Section
\ref{secDiscuss}. Appendices contain the details of our analytical
and numerical calculations.

\section{Polaron spin in a $\pi$- conjugated organic semiconducting
material} \label{SecHamil}

In organic materials, polarons reside on certain molecular or
polymer sites and hop between the sites. While residing on a site
a polaron spin, $\mathbf{S}$, interacts with $N$ surrounding
hydrogen nuclear spins, $\mathbf{I}_j=1/2$, $j=1,..,N$. In a
strong static magnetic field, $\mathbf{B}_0=B_0\hat{\mathbf{z}}$,
this spin dynamics can be described by the Hamiltonian,
\begin{equation}\label{locHam}
H = \Omega S^z + \sum_{j=1}^N S^z\bigl(A_jI^z_j + B_jI^x_j \bigr)
-\omega_I \sum_{j=1}^NI^z_j,
\end{equation}
where $\Omega=\gamma_e\hbar B_0$ and $\omega_I=\gamma_n\hbar B_0$
are the polaron and nuclear Larmor frequencies, and $\{A_j\}$,
$\{B_j\}$ are the hyperfine coupling constants. This
(pseudo)secular description \cite{Slichter} implies that $B_0$
greatly exceeds the local magnetic fields created by the nuclear
magnetic moments, i.e., $\Omega\gg \omega_{\text{hf}}$, where
$\omega_{\text{hf}} = \frac12 \sqrt{\sum_j(A_j^2+B_j^2)}$ is the
average polaron precession frequency in the local field of
surrounding nuclear spins. Assuming measurements in the  X --
band, \cite{BoehmePRL12, Malissa} we will take $B_0\approx 345$~mT
and $\omega_I/2\pi\approx 14.7$~MHz.

The coupling constants in Eq. (\ref{locHam}) depend on the
relative orientation of $\mathbf{B}_0$ and the polaron host
molecular or polymer site. Typically, organic semiconductors are
amorphous materials lacking any long range order in molecular or
polymer orientations. Thus, the coupling constants $\{A_j, B_j\}$
differ from site to site, even if the sites have the same
microscopic structure.
%and averaging over the random molecular or polymer chain
%orientations is often a necessity.

%%%%%%%%%%%%%%%%%%%
\begin{figure}[t]
\centerline{\includegraphics[width=80mm,angle=0,clip]{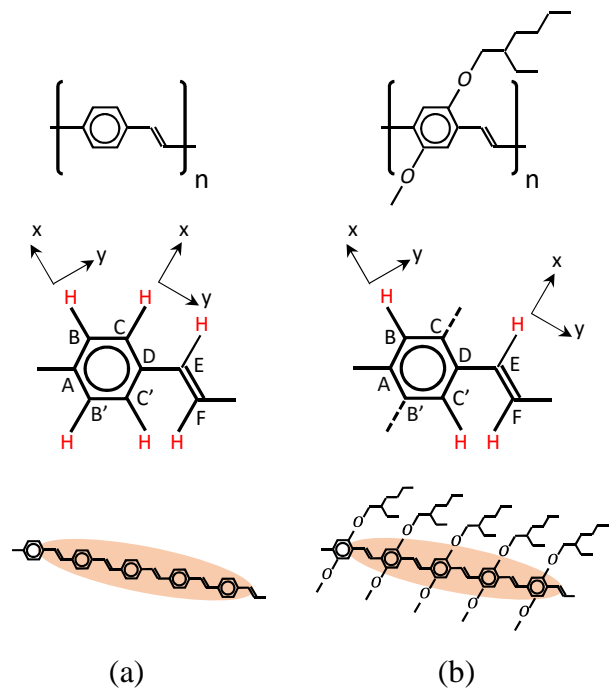}}
\caption{(Color online) Conjugated polymer PPV (a) and its
derivative, MEH-PPV (b). Upper and middle panels: chemical
structures and unit cells. The principal $x$, $y$ axes of the C--H
proton hyperfine tensors at B and C$'$ carbon sites are different
from those at B$'$, C, E, and F carbon sites, while the $z$ axes
are the same and perpendicular to the plane of the picture (in
MEH-PPV there are no C--H protons at B$'$ and C sites). Lower
panels: half-widths of the spatial extents of polarons (brown
ovals), according to Ref.~\onlinecite{Kuroda2000}.} \label{ppv}
\end{figure}
%%%%%%%%%%%%%%%%%%

\subsection*{Polarons in conjugated polymer PPV and MEH-PPV}

The hyperfine interaction between the polaron and proton spins is
determined by the chemical structure of host molecule or polymer,
which also governs the orbital state of the polaron.
%Therefore, in order to be specific, in the following
To be specific, we focus on polymer PPV and its derivative,
MEH-PPV (see Fig.~\ref{ppv}). We base our consideration on the
picture of polaron wavefunction and underlying proton hyperfine
coupling to the polaron spin advocated in Refs.
\onlinecite{Kuroda94PRL, Kuroda95, KurodaSSC95, Kuroda2000}; for a
comprehensive review, see Ref. \onlinecite{Kuroda2003}.

Protons can be naturally divided into two groups. The first group
includes protons located within the envelope of polaron spin
distribution, thus contributing to the {\it contact} hyperfine
interaction. These are all C--H protons covalently coupled to the
polymer backbone carbons, where the polaron wavefunction resides.
Because of an exponentially fast spatial decay the polaron
wavefunction covers a finite number of protons. As it is discussed
below, in PPV and MEH-PPV this number is order of few tens. So
%in our consideration
we neglect contact protons coupled to the polaron spin weaker than
0.5 MHz; the number of such contact protons is small, and their
overall effect is inessential.

Distant protons, which form the second group, couple to a polaron
spin with a magnetic dipolar interaction. These protons belong
both to polymer backbones and substituent side-groups. Simple
estimates show that nearly every distant proton couples to a
polaron spin with less than 1~MHz strength. However, because of
the slow, $\propto 1/r^3$ decay of the dipolar interaction the
effective number of these protons is order of few thousand, so
that their overall effect can be noticeable.

\subsubsection{Contact hyperfine interaction}

The polaron spin, $\mathbf{S}$, couples to a C--H proton spin,
$\mathbf{I}$, by the hyperfine interaction
$\mathbf{S}\cdot\rho_{\text{S}} \hat\mathbf{A}\cdot\mathbf{I}$,
where $\rho_{\text{S}}$ is the polaron spin density on the carbon
$p\pi$ orbital and $\hat\mathbf{A}$ is the hyperfine tensor. Thus
the polaron contact hyperfine interaction is completely described
in terms of $\hat\mathbf{A}$ and $\rho_{\text{S}}$.

From the analysis of unpaired carbon orbital states it is
established \cite{Morton} that the principal $x$ and $z$ axes of
the hyperfine tensor are parallel to the C--H bond and the $p\pi$
orbital axes, respectively (see Fig. \ref{ppv}). Principal
elements of the hyperfine tensor are approximately expressed as
\begin{equation}\label{tensor}
\bigl(A_x, A_y, A_z \bigr) = - \bigl([1-\alpha]A_H, [1+\alpha]A_H,
A_H \bigr),
\end{equation}
where $A_H/2\pi=60$ to $80$ MHz is the McConnell's constant, and
$\alpha= 0.5$ to $0.6$ is the degree of anisotropy. \cite{Morton}

Equation (\ref{tensor}) is quite general for organic
$\pi$-electron radicals. For PPV and MEH-PPV, experimental studies
are conforming with $A_H/2\pi=70$ MHz and $\alpha= 0.5$.
\cite{Kuroda94PRL, Kuroda95, Kuroda2000} These numerical values
are used in our calculations. The remaining necessary ingredient
for describing the polaron contact hyperfine interaction is the
polaron spin density at the carbon sites, $\rho_{\text{S}}$. In
our subsequent calculations we employ the spin density presented
in Table \ref{tab}. This form is found from a model calculation
\cite{KurodaSSC95} and verified by the analysis of spectral
lineshapes in ENDOR \cite{Kuroda94PRL, Kuroda95} and light-induced
ESR \cite{Kuroda2000} experiments.

Formally, $\rho_{\text{S}}$ in Table \ref{tab} is calculated for
PPV. However, the same data can be used for spin densities in
other PPV derivatives, \cite{Kuroda2000} particularly in MEH-PPV,
thus neglecting the effect of substituent groups on
$\rho_{\text{S}}$ to the leading order.

\begin{table}[b]
\caption{Spin density of a polaron in PPV chain,
$\rho_{\text{S}}$,
%inferred
taken from Ref. \onlinecite{KurodaSSC95}. Small values,
$|\rho_{\text{S}}|<0.01$, are
%omitted and subsequently
neglected. The site assignment corresponds to that of Fig.
\ref{ppv}. The unit cell at the polaron center is denoted by $0$,
thereby the unit cells with significant values of
$\rho_{\text{S}}$ range from $-3$ to $3$. }
\begin{tabular}{l |c |c| c| c| c| c|c}
\hline\hline site $\!\!\backslash\!\!$ cell & \,\,\,\,\,
-3\,\,\,\,\,\, & \,\,\,\,\,-2\,\,\,\,\,\, &
\,\,\,\,\,-1\,\,\,\,\,\, & \,\,\,\,\,\, 0\,\,\,\,\,\, &
\,\,\,\,\,\, 1\,\,\,\,\,\,
& \,\,\,\,\,\, 2\,\,\,\,\,\, & \,\,\,\,\,\, 3\,\,\, \\
\hline \,\,$A$  & -- & 0.01 & 0.04 & 0.08 & 0.04 & -- & -- \\
\hline \,\,$B$ & -- & 0.01 & -0.015 & 0.035 & -0.005 & 0.03 & -0.005 \\
\hline \,\,$B'$ & -- & 0.01 & -0.015 & 0.04 & -- & 0.03 & -- \\
\hline \,\,$C$ & -- & -- & 0.03 & -- & 0.04 & -0.015 & 0.01 \\
\hline \,\,$C'$ & -- & -0.005 & 0.03 & -0.005 & 0.035 & -0.015 & 0.01 \\
\hline \,\,$D$ & -- & -- & -- & 0.04 & 0.08 & 0.04 & 0.01 \\
\hline \,\,$E$ & 0.01 & -0.01 & 0.09 & 0.08 & -- & 0.035 & -- \\
\hline \,\,$F$ & -- & 0.035 & -- & 0.08 & 0.09 & -0.01 & 0.01 \\
\hline
\end{tabular}
\label{tab}
\end{table}

%In what follows we
We further restrict ourselves on MEH-PPV. According to Table
\ref{tab} and Fig.~\ref{ppv}, in MEH-PPV there are $N_c=22$
contact proton spins coupled to a polaron spin at sites $B$, $C'$,
$E$, and $F$, over 7 consecutive unit cells that the polaron spin
distribution covers (note that in MEH-PPV the C--H protons at
carbon sites B and C$'$ are replaced by substituent groups). In
the Hamiltonian (\ref{locHam}) we label the contact protons by
$j=1,..,N_c$. The coupling constants $\{A_j, B_j\}_{j=1}^{N_c}$
depend on the relative orientations of the corresponding C--H
bonds and the applied magnetic field, $\mathbf{B}_0 =
B_0\hat{\mathbf{z}}$. We denote the components of
$\hat{\mathbf{z}}$ in the principal basis of the $j$- th hyperfine
tensor by $q_{\mu j}$, $\mu=x$, $y$, $z$. The coupling constants
are related to the hyperfine tensor elements Eq. (\ref{tensor}) as
\begin{equation}\label{ABc}
A_j= \rho_{\text{S}}(j)\! \sum_\mu A_\mu q_{\mu j}^2, \quad A_j^2
+ B_j^2 = \rho_{\text{S}}^2(j)\! \sum_\mu A_\mu^2q_{\mu j}^2.
\end{equation}
For each $j$, $\rho_{\text{S}}(j)$ is given in Table \ref{tab},
and $q_{\mu j}$ can be found for any direction of $\mathbf{B}_0$
from the description of principal hyperfine axes in Fig.
\ref{ppv}. Thus finding the coupling constants and performing the
orientation averaging of different quantities of interest becomes
a straightforward numerical task. As the first step we calculate
the average local frequency due to the contact hyperfine coupling,
\begin{equation}\label{OMc}
\omega_{\text{hf},c} = \left\langle \frac12 \sqrt{\sum_{j\leq
N_c}\bigl( A_j^2 +B_j^2 \bigr)} \right\rangle \approx 2\pi \times
7.25\, \text{MHz}.
\end{equation}
The corresponding ESR line would be Gaussian, with the full width
at half maximum of $6.1$~G, in agreement with
Ref.~\onlinecite{Kuroda2000}.

\subsubsection{Interaction with the distant protons}

Distant protons couple to the polaron local spin density via
magnetic dipolar interaction. The strength of this interaction is
determined by the material morphology, including the molecular
packing and the average density of protons. Relying upon the
reported data on the molecular packing \cite{Claes01, Kilina13,
Qin13} and van der Waals radii of hydrogen and carbon \cite
{Bondi64, Motoc85, Spillane92}, we restrict the minimal distance
between the polymer backbone carbons and distant protons to
$d_{\text{min}} = 2.2$~\AA. Furthermore, based on the MEH-PPV mass
density, $1$~g/mL, \cite{Kilina13, Qin13} and its chemical
structure in Fig.~\ref{ppv}, we infer the average proton density,
55~nm$^{-3}$. Except in the regions restricted by $d_{\text{min}}$
around a polymer chain, we take random distributions of protons
with this average density. We also employ point dipolar coupling
between a distant proton and each of the 38 non-zero polaron spin
densities, given in Table~\ref{tab}.
%According to the Table, there are 34 carbon sites carrying a
%non-zero polaron spin density.
Thus, in the applied magnetic field, $\mathbf{B}_0 =
B_0\hat{\mathbf{z}}$, a distant proton couples to the polaron spin
with the Hamiltonian (\ref{locHam}), where
\begin{eqnarray}\label{ABd}
&&A= \hbar\, \gamma_e \gamma_n
\sum_{l=1}^{38}\rho_{\text{S}}(l)\frac{1- 3\cos^2\theta_l}
{R_l^3}, \nonumber \\
&&B= \hbar\, \gamma_e \gamma_n \sum_{l=1}^{38}\rho_{\text{S}}(l)
\frac{3\sin\theta_l \cos\theta_l}{R_l^3}.
\end{eqnarray}
Here, $\rho_{\text{S}}(l)$ is the polaron spin density at the
carbon site $l$, $\mathbf{R}_l$ is the vector connecting the
distant proton to this carbon site, and $\theta_l$ is the angle
between $\mathbf{R}_l$ and $\mathbf{B}_0$.

Equations (\ref{ABd}) give coupling constants of a single distant
proton. A large number of distant protons is included numerically,
by sampling realizations of their spacial distributions. In our
simulations of various quantities of interest the results converge
for about $N_d=2000$ uniformly distributed distant protons and do
not change appreciably if this number is increased by an order of
magnitude. This is because we deal with spatial integrals of $\sim
A^2, B^2$, and their combinations, which vanish $\propto R^{-6}$
or faster, and thus converge quickly. Averaging over the random
orientations of polymer chains should be performed additionally,
as the polaron spin density is not spherically symmetric and
different chain orientations are inequivalent.

For the average local frequency from the distant protons this
yields $\omega_{\text{hf},d}\approx 2\pi \times 2$~MHz, leading to
the the total average local frequency,
\begin{equation}\label{OMt}
\omega_{\text{hf}} = \left\langle \frac12 \sqrt{\sum_{\text{all }
j}\bigl( A_j^2 +B_j^2 \bigr)} \right\rangle \approx 2\pi \times
7.52\, \text{MHz}.
\end{equation}
%Equations (\ref{OMc}) and (\ref{OMt}) give the total average local
%frequency, $\omega_{\text{hf}}= 2\pi \times 7.52$. Note that
From Eqs. (\ref{OMc}) and (\ref{OMt}) it is seen that, on average,
the distant protons are responsible only for a small fraction of
the local hyperfine field. Yet they have a strong effect on the
fine structure of ESEEM, as will be seen shortly.

In theoretical studies of spin dynamics in organic semiconductors
a semiclassical approach \cite{SchulWol} is often used. While this
approach does not capture the fine structure of the ESEEM signal,
it provides a convenient way for the characterization of the
signal decay. In the semiclassical treatment, the nuclear spin
dynamics given by the last term of Eq. (\ref{locHam}) is  ignored,
and the on-site hyperfine interaction is replaced by a random
local static magnetic field felt by a polaron spin.
\cite{SchulWol} Accordingly, the on-site semiclassical Hamiltonian
in the secular approximation reads:
\begin{equation}\label{QC}
H_{SC} = \bigl(\Omega +\omega_z) S^z,
\end{equation}
where $\omega_z$ is random and uncorrelated from site to site.
This random frequency is often described by a Gaussian
distribution. In the case under consideration the Gaussian
distribution of frequencies should be taken with the standard
deviation, $\omega_{\text{hf}}$. Note that the distribution of
random fields resulting from a magnetic dipolar bath of randomly
spaced spins is rather Lorentzian, with a certain frequency
cutoff. \cite{Abragam} In our case this is pertinent to the
contribution of distant protons. However, a relatively short
cutoff and overall small contribution of distant protons in
$\omega_{\text{hf}}$ ensure that the Gaussian distribution of
local frequencies is accurate also in our case.

\section{Spin echo with ideal pulses}
\label{SecDeriv}

\begin{figure}[t]
\centerline{\includegraphics[width=80mm,angle=0,clip]{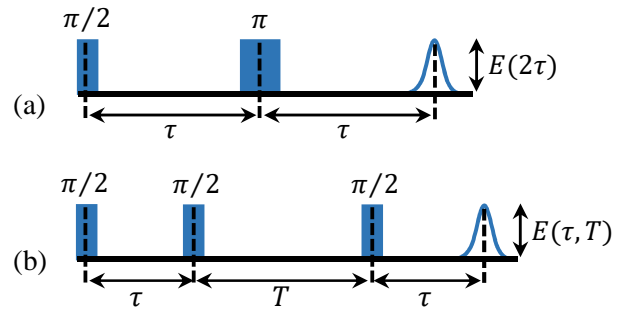}}
\caption{(Color online) ESEEM pulse sequences considered in the
text. (a) Primary ESEEM. (b) Stimulated ESEEM. } \label{seqs}
\end{figure}

Generally, electron spin echo envelope modulation (ESEEM)
spectroscopy is used to investigate the hyperfine interactions of
paramagnetic species. \cite {DT} To set a framework for discussing
the application of spin echo experiments in organic semiconductors
we give a treatment of the ESEEM originating from the two-pulse,
Hahn echo sequence, Fig. \ref{seqs}(a)  (primary ESEEM), and the
three-pulse sequence, Fig. \ref{seqs}(b) (stimulated ESEEM). In
Fig. \ref{seqs}, $\pi/2$ and $\pi$ denote the rotation angle of
spins along the $x$-axis in the rotating frame, induced by
resonant microwave pulses, whereas $\tau$ and $T$ are the free
evolution periods between the pulses. The pulses are assumed to be
ideal. Depending on $\tau$ and $T$ the echo amplitude undergoes
modulations, which we denote by $E(2\tau)$ for the primary ESEEM
and $E(\tau,T)$ for the stimulated ESEEM.

%%%%%%%%%%%%%%%%%%%
%%%%%%%%%%%%%%%%%%

Using the density matrix formalism, the (normalized) modulation
functions can be written as
\begin{eqnarray}\label{eef}
&&E(2\tau)= -2\, \text{Tr}\left[U(\tau) \hat{\rho}(0)
U^\dag(\tau) S^y\right],\\
&&E(\tau,T)= -2\, \text{Tr}\left[U(\tau, T) \hat{\rho}(0)
U^\dag(\tau, T) S^y\right],
\end{eqnarray}
where $\hat{\rho} (0)$ is the density operator before the first
pulse, and the evolution operators  are given by
\begin{eqnarray}\label{evop}
&&U(\tau)= e^{-i\tau H}\bigl[\pi\bigr]
e^{-i\tau H}\bigl[ \pi/ 2\bigr], \nonumber \\
&&U(\tau,T)=  e^{-i\tau H}\bigl[ \pi/ 2\bigr] e^{-iT
H}\bigl[\pi/2\bigr] e^{-i\tau H}\bigl[ \pi/ 2\bigr], \nonumber
\end{eqnarray}
in which $[\phi]=\exp( i\phi S^x)$ is the rotation operator for an
ideal pulse with flip angle $\phi$, and $H$ is the Hamiltonian,
Eq. (\ref{locHam}). For our further purposes we consider the
initial density operator, $\hat {\rho}(0)= (1/2 +
S^z)\otimes\rho_I$, formally describing a polaron spin ensemble
polarized along the $z$-axis. At the same time we neglect the
thermally-induced polarization of the nuclear spin ensemble and
take the nuclear density operator proportional to the unity,
$\rho_I \propto\mathbf{1}$. The explicit calculation of modulation
functions is facilitated by the fact that the Hamiltonian, Eq.
(\ref{locHam}), preserves the $z$-component of polaron spin. One
gets \cite{DT}
\begin{equation}\label{ndprE}
E(2\tau)= \prod_{j=1}^N\!\!\left(1-2k_j\sin^2\frac{\omega_{j
+}\tau}2 \sin^2 \frac{\omega_{j -}\tau}2 \right)
\end{equation}
for the primary ESEEM and
\begin{eqnarray}\label{ndstE}
E(\tau,T)= \frac 12 \prod_{j=1}^N\!\!\left(1-2k_j\sin^2
\frac{\omega_{j +}[\tau+T]}2 \sin^2 \frac{\omega_{j -}\tau}2
\right)&& \nonumber \\
+\frac 12 \prod_{j=1}^N\!\!\left( 1- 2k_j\sin^2 \frac{\omega_{j +}
\tau}2 \sin^2 \frac{\omega_{j -}[\tau+T]}2 \right)\quad &&
\end{eqnarray}
for the stimulated ESEEM, where the frequencies,
\begin{equation}\label{endorfq}
\omega_{j\pm}= \left[(\omega_I \pm A_j/2)^2 +B_j^2/4
\right]^{1/2},
\end{equation}
are the nuclear spin level splittings due to the external magnetic
field and the hyperfine coupling to the polaron spin in either up
or down state, and
\begin{equation}\label{moddepth}
k_j=\left[\frac{\omega_I B_j}{\omega_{j +} \omega_{ j -}}\right]^2
\end{equation}
are the modulation depths. The primary ESEEM formula (\ref{ndprE})
is a product of the factors from individual nuclei. This is the
consequence of the polaron spins being in a coherent superposition
of up and down states during the evolution periods. In the
stimulated ESEEM, during the evolution time $T$ the polaron spins
are in a definite state, up or down, leading to the first and the
second products in Eq. (\ref{ndstE}), respectively.

Two major factors influencing modulation signals Eqs.
(\ref{ndprE}) and (\ref{ndstE}) in a real experiment are the
orientation disorder of polymer chains and the polaron random
hopping between the sites. In the next two Sections we address the
effects of these factors.

\section{The effect of orientation disorder}
\label{SecOrDis}

The samples in typical experiments on organic polymers are
disordered films, and the observed signals incorporate
contributions from all orientations of
%randomly distributed
polymer chains. Therefore we average Eqs. (\ref{ndprE}) and
(\ref{ndstE}) over random orientations of polymer chains, and
consider the disorder-averaged modulation signals, $\langle
E(2\tau) \rangle$, $\langle E(\tau,T) \rangle$, together with
their spectra given by the cosine Fourier transforms, \cite{ftnt}
$\tilde{E}(\omega) = \mathcal{F}_\tau[ \langle E(2\tau) \rangle]$,
$\tilde{E}(\tau, \omega) = \mathcal{F}_T [ \langle E(\tau,T)
\rangle]$.

\subsection{Orientation-averaged primary ESEEM}

%%%%%%%%%%%%%%%%%%%
\begin{figure}[t]
\vspace{-0.2cm}
\centerline{\includegraphics[width=95mm,angle=0,clip]{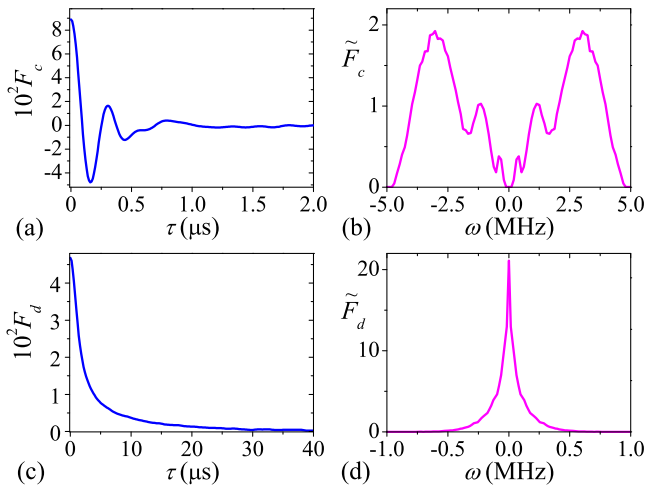}}
\vspace{-0.4cm} \caption{(Color online) Functions $F_c(\tau)$ and
$F_d(\tau)$, introduced in Eqs. (\ref{Fc}) and (\ref{Fd})
respectively, are plotted with blue. (b) and (d): The respective
cosine Fourier transforms, $\tilde{F}_c(\omega)$ and
$\tilde{F}_d(\omega)$, are plotted with magenta, in the same
units.  } \label{fcfd}
\end{figure}
%%%%%%%%%%%%%%%%%%

The HFI described above leads to small values of modulation
depths, $k_j\ll 1$. Moreover, the sum of all modulations depths,
$\kappa = \sum k_j$, is also small. This alludes to expanding Eq.
(\ref{ndprE}) over small $k_j$ (for more details on this approach
see Appendix \ref{AppA}). We write:
\begin{eqnarray}\label{aprprE}
E(2\tau)=1- \frac 12\sum_j k_j \biggl[1 -\cos(\omega_{j +}\tau)
-\cos(\omega_{j -}\tau)&&  \nonumber \\
\left . + \frac 12\cos\bigl( [\omega_{j +} - \omega_{j -}]\tau
\bigr) + \frac 12 \cos\bigl( [\omega_{j +} + \omega_{j -} ] \tau
\bigr)\right ]\! .&& \quad
\end{eqnarray}
Equation (\ref{aprprE}) shows that the primary ESEEM spectrum
involves four groups of carrier frequencies, $\{\omega_{j \pm}\}$
and $\{\omega_{j +} \pm \omega_{j -}\}$. We subsequently use the
approximation,
\begin{equation}\label{ompmapr}
\omega_{j \pm} \approx \omega_I \pm A_j/2.
\end{equation}
For the distant protons Eq. (\ref{ompmapr}) follows from the weak
coupling, $A_j, B_j \ll \omega_I$. For the contact protons with a
stronger coupling Eq. (\ref{ompmapr}) is valid due to the weak
anisotropy of the contact HFI, see Appendix \ref{AppA}. Employing
Eq. (\ref{ompmapr}) the four frequency groups involved are
$\{|A_j|\}$, $\{ \omega_I - |A_j|/2\}$, $\{ \omega_I + |A_j|/2\}$,
and $2\omega_I$. The relation, \cite{ftnt2} $\omega_I>\frac 32
|A_j|$, ensures that $\{|A_j|\}$ carries the low-frequency
modulations, resolved from the higher frequency groups. Besides,
the second and the third groups are close to $\omega_I$, mirroring
each other about this frequency.

Another conclusion from Eq. (\ref{aprprE}) is that the
%Also,
contributions of the contact and the distant protons in $E(2\tau)$
are simply additive. We separate these contributions by
introducing the notations, $E_c(2\tau)$ and $E_d(2\tau)$,
respectively. More specifically, $E_c(2\tau)$ is the partial sum
of the first $N_c$ terms in Eq.~(\ref{aprprE}), whereas
$E_d(2\tau)$ is that of the terms with $j>N_c$, and thus $E(2\tau)
= 1 + E_c(2\tau) + E_d(2\tau)$. Using Eq. (\ref{ompmapr})
%the above expansion of $\omega_{j \pm}$
in Eq. (\ref{aprprE}) and averaging the result over the disorder
in polymer chain orientations yields:
\begin{eqnarray}\label{Etot}
\langle E_\beta (2\tau) \rangle = - \frac{\langle \kappa_\beta
\rangle}2 - \frac 14F_\beta (2\tau) - \frac{\langle \kappa_\beta
\rangle}4 \cos (2\omega_I \tau)&& \nonumber \\
+ F_\beta (\tau)\cos (\omega_I \tau),&&
\end{eqnarray}
where the subscript, $\beta=c, d$, refers to the contact and the
distant protons, respectively, and the partial sums
\begin{eqnarray}\label{Fc}
&& F_c(\tau)= \left\langle \sum\nolimits_{j\leq N_c} k_j\cos
(A_j\tau/2) \right\rangle, \\
&& F_d(\tau)= \left \langle \sum\nolimits_{j> N_c} k_j\cos
(A_j\tau/2) \right \rangle, \label{Fd}
\end{eqnarray}
$\kappa_c =\langle\sum_{j=1}^{N_c} k_j \rangle $,
%\approx 0.089$,
and $\kappa_d =\langle\sum_{j>N_c} k_j\rangle$
%\approx 0.047$ are found numerically,
are introduced. Equation (\ref{Etot}) gives the
orientation-averaged ESE modulation function in terms of
$F_c(\tau)$ and $F_d(\tau)$. Particularly, the low-frequency
modulations are  included in the second term of Eq. (\ref{Etot}).
The third term of Eq. (\ref{Etot}) describes oscillations of a
constant amplitude on the frequency $2\omega_I$, both for the
contact and the distant protons. Finally, modulations with the
frequencies close to $\omega_I$ are incorporated in the last term
of Eq. (\ref{Etot}).

%providing the most useful information on the primary ESEEM, and as
%will be seen shortly, allowing a clear resolution of the contact
%and the distant proton contributions.
%Its cosine Fourier transforms is simply $\tilde{F}(\omega -
%\omega_I)$, where $\tilde{F}(\omega) = \mathcal{F}_\tau[ F(\tau)]$.

First we elaborate on the contribution of distant protons.
%$\langle E_d(2\tau) \rangle$.
On the timescale, $\tau\sim 1/\omega_I$, the function $F_d(\tau)$,
Fig.~\ref{fcfd}(c), varies only slightly. Thus the last term of
$\langle E_d(2\tau) \rangle$ in Eq. (\ref{Etot}) can be viewed as
oscillations with the frequency, $\omega_I$, and the envelope,
$F_d(\tau)$. The cosine Fourier transform, $\tilde{F}_d(\omega)$,
plotted in Fig.~\ref{fcfd}(d), is a sharp peak at $\omega=0$.
% nearly vanishing for $|\omega|>0.5$~MHz.
Through this function
%and Eq. (\ref{Etot})
the cosine Fourier spectrum of the distant protons
%, $\tilde{E}_d(\omega)$,
is found. It involves three well-resolved peaks; a dip of the
form, $ -\frac14 \tilde{F}_d(\omega/2)$, near the origin, a sharp
peak at $\omega_I$ of the shape, $\tilde{F}_d(\omega-\omega_I)$,
and a sharper, negative $\delta$- peak at $2\omega_I$.

In the case of the contact proton contribution, the function
$F_c(\tau)$, Fig. \ref{fcfd}(a), changes considerably on the
timescale, $\tau\sim 1/\omega_I$, because of the presence of large
$A_j\sim \omega_I$. Therefore the last term of $\langle E_c(2\tau)
\rangle$ in Eq. (\ref{Etot})
%the time domain signal $\langle E_c(2\tau)\rangle$
does not admit a simple interpretation in terms of the
oscillations with the frequency $\omega_I$ and a smooth envelope.
%As seen in Fig.~\ref{fcfd}(b), the cosine Fourier transform
%$\tilde{F}_c(\omega)$ consists of two bands mirroring each other
%about $\omega=0$.
%Still,
Its cosine Fourier transform,
% is simply
$\tilde{F}_c(\omega -\omega_I)$, incorporates two bands mirroring
each other about $\omega_I$, as can be inferred from
Fig.~\ref{fcfd}(b). These bands come from the modes with
frequencies, $\{\omega_I \pm |A|_j/2 \}_{j=1}^{N_c}$, spread by
the orientation disorder. Except for these two bands and the
negative $\delta$- peak at $2\omega_I$, the cosine Fourier
spectrum of the contact protons
%$\tilde{E}_c(\omega)$,
involves a low-frequency band of the form,
$-\frac14\tilde{F}_c(\omega/2)$, originating from the frequencies
$\{|A|_j\}_{j=1}^{N_c}$.

\begin{figure}[t]
\vspace{-0.3cm}
\centerline{\includegraphics[width=95mm,angle=0,clip]{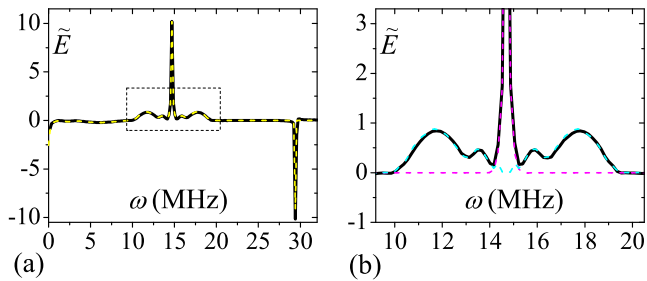}}
\vspace{-0.4cm} \caption{(Color online) The primary ESEEM spectrum
$\tilde{E}$, calculated from orientation disorder averaged Eq.
(\ref{ndprE}) numerically, is plotted in black. (a): The cosine
Fourier transform of the sum, $\langle E_c(2\tau) \rangle +
\langle E_d(2\tau) \rangle$, is plotted with yellow dashed line,
from  Eq. (\ref{Etot}). (b): Zoom in of the region indicated in
the left panel with a rectangle. $\tilde{F}_c(\omega - \omega_I)$
and $\tilde{F}_d(\omega - \omega_I)$ are plotted with the cyan and
magenta dotted lines, respectively. It is seen that the spectral
peak at $\omega_I=14.7$~MHz is exclusively due to the distant
protons, whereas the side bands come from the contact protons.}
\label{primE}
\end{figure}

Figure~\ref{primE} plots the primary ESEEM spectrum,
$\tilde{E}(\omega)$, calculated from Eq. (\ref{ndprE}) by a Monte
Carlo sampling of the polymer chain orientations, employing Eqs.
(\ref{ABc}),~(\ref{ABd}). Its structure near $\omega_I= 14.7$~MHz
%Fig.~\ref{primE}(b),
includes a sharp peak at $\omega_I$ and two
wider side-bands mirroring each other about $\omega_I$. Based on
the above analysis we identify the side-bands with the
contribution of contact protons and the sharp peak with that of
the distant protons. Thus, the forms of the side-bands and the
sharp peak are given by $\tilde{F}_c(\omega - \omega_I)$ and
$\tilde{F}_d(\omega - \omega_I)$, respectively.
%which add up to $\tilde{F}(\omega - \omega_I)$.
This identification is clearly confirmed in Fig. \ref{primE}(b)
where we plot the contributions of contact and distant protons
separately.

%%%%%%%%%%%%%%%%%%%
%%%%%%%%%%%%%%%%%%

\subsection{Orientation-averaged stimulated ESEEM}

The stimulated ESEEM can be analyzed along the same lines.
Expanding Eq. (\ref{ndstE}) over small $k_j$ and keeping the
leading terms, one gets:
\begin{eqnarray}\label{steapr}
E (\tau,T) = 1 - \sum_j\frac{k_j}2 \Bigl [\sin^2 \frac{\omega_{j
+}\tau}2 \bigl( 1- \cos\omega_{j -} [\tau+T] \bigr) \nonumber \\
+ \sin^2\frac{\omega_{j -}\tau}2\bigl( 1- \cos \omega_{j
+}[\tau+T]\bigr) \Bigr].\qquad&&
\end{eqnarray}
%This indicates that
Thus the stimulated ESEEM spectrum involves only two groups of
frequencies, $\{\omega_{j+}\}$ and $\{\omega_{j-}\}$. Our
subsequent analysis employs the approximation Eq. (\ref{ompmapr}).
By separating the contact and the distant proton contributions in
Eq. (\ref{steapr}) and averaging over the
%random
polymer chain orientations we get $\langle E (\tau,T) \rangle = 1
+ \langle E_c (\tau,T)\rangle + \langle E_d (\tau,T)\rangle$,
where the $T$- dependent parts
%of the contact and the distant proton contributions have the form,
of $\langle E_\beta (\tau,T)\rangle$, $\beta = c, d$, are
%given by
\begin{eqnarray}
\label{Tpart} &&\langle E_\beta (\tau,T) \rangle \simeq \frac 12
F_\beta(\tau + T) \cos (\omega_I [\tau + T]) \\
&&-\frac 14 F_\beta(T) \cos (\omega_I [2\tau+T] )  -\frac 14
F_\beta(2\tau + T) \cos (\omega_I T). \nonumber
\end{eqnarray}
%This equation defines the $T$- dependence of $\langle E (\tau,T)
%\rangle$ in terms of the known functions.

As a function of $T$, $\langle E_d (\tau,T)\rangle$ modulates only
with the proton Zeeman frequency, $\omega_I$, and its cosine
Fourier transform, \cite{ftnt} $\tilde{E}_d (\tau,\omega)$, is
basically a sharp peak around that frequency. Predictions about
the $\tau$- dependence of the modulation strength can be made even
without performing the disorder averaging. Indeed, Eq.
(\ref{steapr}) shows that the modulation amplitude is reduced if
$\tau$- values are possible such that $\sin(\omega_{j
\pm}\tau/2)\approx 0$ for all protons. Since for the distant
protons all $\omega_{j \pm}$ are close to $\omega_I$, one can
expect a reduction of the modulation amplitude of $\langle E_d
(\tau,T)\rangle$ at the $\tau$- values with
$\sin(\omega_I\tau/2)=0$. Similarly, one can anticipate an
increase of modulation amplitude at the $\tau$- values with
$\sin(\omega_I\tau/2)=\pm 1$.

%%%%%%%%%%%%%%%%%%%
\begin{figure}[t]
\vspace{-0.3cm}
\centerline{\includegraphics[width=95mm,angle=0,clip]{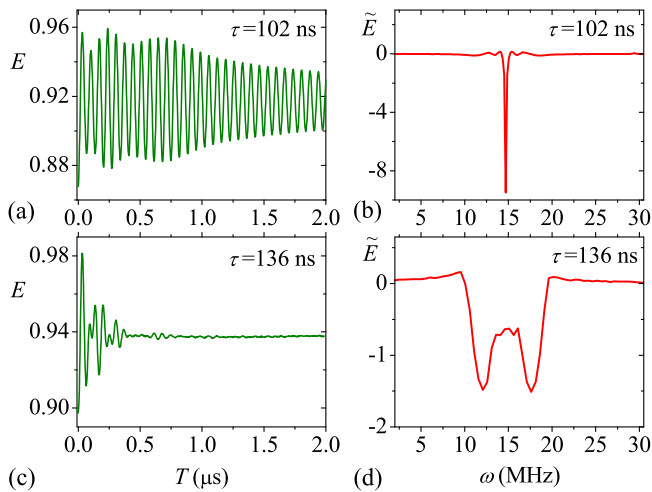}}
\vspace{-0.4cm} \caption{(Color online) The stimulated ESEEM
$\langle E(\tau_n,T)\rangle$, calculated from orientation disorder
averaged Eq. (\ref{ndstE}) numerically, is plotted against $T$ at
fixed $\tau_n=(\pi/\omega_I)n$ for $n=3,\, \tau_3 \approx 102$~ns
(a), and $n=4,\, \tau_4 \approx 136$~ns (c). The corresponding
spectra $\tilde{E}(\tau_n,\omega)$, $n=3$ (b) and $n=4$ (d), are
plotted against $\omega$. The strong reduction of the peak at
$\omega_I=14.7$~MHz for even $n$, allowing the observation of the
contact proton hyperfine coupling, is obvious.} \label{efct0}
\end{figure}
%%%%%%%%%%%%%%%%%%

In Appendix \ref{AppA} we prove that the $T$- modulation amplitude
of $\langle E_d (\tau,T)\rangle$ is reduced at the $\tau$- values,
$\tau_n=(\pi/\omega_I)n$ with even integer $n$, and increased at
those with odd integer $n$. We also show that, for $n\leq 30$, the
decrease in the amplitude of $\langle E_d (\tau_n,T)\rangle$
between odd $n$ and even $n$ is more than two orders of magnitude
for small $n$ and more than 15 times for large $n$. Note that this
includes all $\tau_n$ within the interval, $0<\tau<1\,\mu$s, which
covers the experimentally available $\tau$- domain, taking into
account the decay of the signal in a real experiment.

The $T$- modulation of $\langle E_c (\tau,T)\rangle$, given by Eq.
(\ref{Tpart}), cannot be interpreted as having a single frequency
$\omega_I$, because the function $F_c$ varies abruptly on the
timescale, $T\sim 1/\omega_I$. Similar to that of the primary
modulation, its cosine Fourier transform, \cite{ftnt} $\tilde{E}_c
(\tau,\omega)$, represents two bands near $\omega_I$. However, in
this case these bands are not quite symmetric around $\omega_I$.
Importantly, $\tau = \tau_n$ are not critical for $\langle E_c
(\tau,T)\rangle$ and there is no special reduction at even $n$, as
it is shown in Appendix \ref{AppA}.

Summarizing, measurements of the stimulated ESEEM spectra at $\tau
= \tau_n$ with odd $n$ encounter a strong peak at $\omega_I$,
which in a real situation could make the observation of weaker
contact proton sidebands difficult. On the other hand, reduction
of the peak occurs at $\tau = \tau_n$, while the contribution of
the contact protons is preserved. This constitutes a basis of the
method for addressing the distant and the contact hyperfine
protons selectively, by choosing appropriate $\tau$- values and
analyzing the $T$-dependence of the stimulated ESEEM.

%Throughout our simulations we see this effect consistently.
To illustrate the method, in Fig.~\ref{efct0} we plot the
time-domain signals $\langle E(\tau_n,T)\rangle$ and their spectra
for $n=3$ and $n=4$, calculated from orientation-averaged exact
equation (\ref{ndstE}) numerically.
%by Monte Carlo sampling the polymer chain orientations.
Spectral plots in Figs.~\ref{efct0}(b) and (d) demonstrate the
suppression of the peak at $\omega_I=14.7$~MHz when changing $n$
from odd to even.

\section{Echo modulations of hopping polarons}
\label{SecHopping}

The  polaron random walk leads to the
%dephasing
decay of ESEEM,
%and its subsequent decay,
thus imposing limitations on the observability of modulations. On
the other hand, this decay can serve as a probe for understanding
the aspects of polaron transport. In this Section we investigate
the ESEEM of polarons performing random walk over orientationally
disordered polymer sites and coupling to the nuclear spins
according to Eq. (\ref{locHam}). Our main goal is to reveal the
hopping regimes where the ESEEM signal,
%is amenable for a spectral analysis, and particularly, where
and particularly the contact hyperfine spectrum, is not distorted.
%Our main goal is to reveal the hopping regimes where the ESEEM is
%observable, and its spectrum unambiguous.

The spin dynamics of a randomly hopping polaron is dependent on
the polaron random walk dimensionality. \cite{MD15, RR14, CzK,
MLD} Its analytical description is the simplest in 3D, where the
approximation neglecting the self-intersections of the random walk
trajectories is good. This is equivalent to the strong collision
approximation which provides a simple way of describing the spin
relaxation of a randomly hopping carrier. \cite{Kubo79}

The multiple trapping model \cite{Hartenstein96, Jakobs93,
HarmonPRL13, HarmonPRB14} is an implementation of the strong
collision approximation, often used to explain the transport in
organic materials \cite{Coropceanu07} and particularly in PPV and
its derivatives. \cite{Blom00} We base our consideration on the
multiple trapping model. Within this model the polaron hopping
from a polymer site, $\mathbf {r}$, is described by the rate,
\begin{equation}\label{MTM}
W_{\mathbf {r}} = \nu\exp\bigl[\varepsilon_{\mathbf {r}}/k_BT
\bigr],
\end{equation}
where $\nu$ is the hopping attempt frequency, $\varepsilon_
{\mathbf {r}}$ is the trapping energy at the site $\mathbf {r}$,
$k_B$ is the Boltzmann constant, and $T$ is the temperature. The
trapping energies are all negative and random, with the
exponential distribution, $\mathcal{N} (\varepsilon)\propto\exp
\bigl[\varepsilon/ k_B T_0 \bigr]$. Hence the model is defined by
two parameters, $\nu$ and the dispersion parameter, $\alpha \equiv
T/T_0$. In the high-temperature or shallow-trap limit, $\alpha \to
\infty$, the hopping rates are uniform and the waiting time
statistics of the polaron random walk is governed by the Poisson
distribution, $P(t)=\nu\exp(-\nu t)$. For finite $\alpha$ this
distribution assumes the algebraic form, $P( t)\propto
t^{-1-\alpha}$, reflecting the broad distribution of hopping
rates.

\subsection{Primary ESEEM of hopping polarons}

The generalization of Eq.~(\ref{ndprE}) for hopping polarons and
%the numerical procedure for
the evaluation of the resulting echo modulation function,
$\mathcal{E}(2\tau)$, is described in Appendix \ref{AppC}. We
calculate $\mathcal{E}(2\tau)$ by Monte-Carlo sampling of the
random-walk trajectories over the orientation disordered polymer
sites.
%The results on $\mathcal{E}(2\tau)$ are conveniently given in terms of
But before turning to our results on $\mathcal{E}(2\tau)$ we
introduce the echo modulation function of hopping carriers
calculated from the semiclassical Hamiltonian~(\ref{QC}),
$\mathcal{E}_{\text{SC}}(2\tau)$, which is the semiclassical
counterpart of $\mathcal{E}(2\tau)$.

$\mathcal{E}_{\text{SC}}(2\tau)$ is a non-oscillatory,
monotonously decreasing function of the delay time $\tau$. In the
high-temperature limit, $\alpha \to \infty$, the perturbative
treatment over small $\eta \equiv \nu/ \omega_{\text{hf}} \ll 1$
given in Appendix \ref{AppB} yields
\begin{equation}\label{slowsc}
\mathcal{E}_{\text{SC}}(2\tau)= \left[ 1+ \eta\sqrt{\pi}\,
\text{erf}(\omega_{\text{hf}}\,\tau ) \right]\! e^{-2\nu\tau},
\end{equation}
where $\text{erf}(x)$ is the error function. For
$\tau>2/\omega_{\text{hf}}$ the error function in Eq.
(\ref{slowsc}) changes very little, so that
$\mathcal{E}_{\text{SC}}(2\tau)$ assumes the exponential form,
$\mathcal{E}_{\text{SC}}(2\tau) \propto\exp(-2\tau/T_2)$, with the
decoherence time, $T_2=1/\nu$. The decay of
$\mathcal{E}_{\text{SC}} (2\tau)$ with $\tau$ is exponential also
in the fast hopping regime, $\eta\gg1$. However, due to the
motional narrowing, the dependence of $T_2$ on $\nu$ in this
regime is reversed; $T_2 = \nu/ \omega_{\text{hf}}^2$. Combining
the two
%regimes, one can write:
forms, we write:
\begin{equation}\label{T2}
T_2=1/\nu +\nu/\omega_{\text{hf}}^2.
\end{equation}
Even though the decay of $\mathcal{E}_{\text{SC}}(2\tau)$ in the
intermediate regime $\eta\sim 1$ is not exponential, the dephasing
time Eq.~(\ref{T2}) gives the correct timescale for that decay
too.

Our numerical simulations show that with decreasing $\alpha$ the
decay of $\mathcal{E}_{\text{SC}} (2\tau)$ becomes slower and
non-exponential, with a progressively stronger long-time tail. For
$\eta\ll 1$ this can be explained as follows. The dependence of
$\mathcal{E}_{\text{SC}} (2\tau)$ on $\alpha$ is stipulated by the
number of deep traps, which grows with decreasing $\alpha$. A
trapped polaron is subject to a static hyperfine magnetic field.
Because the echo pulse sequence eliminates the dephasing caused by
static field components, \cite{Abragam, Slichter} the decay of
$\mathcal{E}_{\text{SC}} (2\tau)$ becomes slower with the
increasing fraction of trapped polarons. The effect is most
pronounced at long times due to the slow, algebraic decrease of
the waiting time distribution, resulting in the overall
non-exponential dephasing of $\mathcal{E}_{\text{SC}} (2\tau)$.
%The slower decay for $\eta\lesssim 1$ is also consistent with the
%fact that with decreasing $\alpha$ the mean hopping rate is
%reduced.

The dependence of $\mathcal{E}_{\text{SC}} (2\tau)$ on $\alpha$
for $\eta \gg 1$ is less transparent. Nevertheless, the
non-exponential character of $\mathcal{E}_{\text{SC}} (2\tau)$ at
finite $\alpha$, observed in our numerical simulations, is
established analytically also for this case. \cite{YMR}

Summarizing, the exponential behavior of $\mathcal{E}_{\text{SC}}
(2\tau)$ is a signature of the uniform hopping rates with either
fast or slow hopping (i.e., away from $\eta \sim 1$), whereas in
all the remaining situations $\mathcal{E}_{\text{SC}} (2\tau)$ is
non-exponential.

%%%%%%%%%%%%%%%%%%%
\begin{figure}[t]
\vspace{-0.4cm}
\centerline{\includegraphics[width=95mm,angle=0,clip]{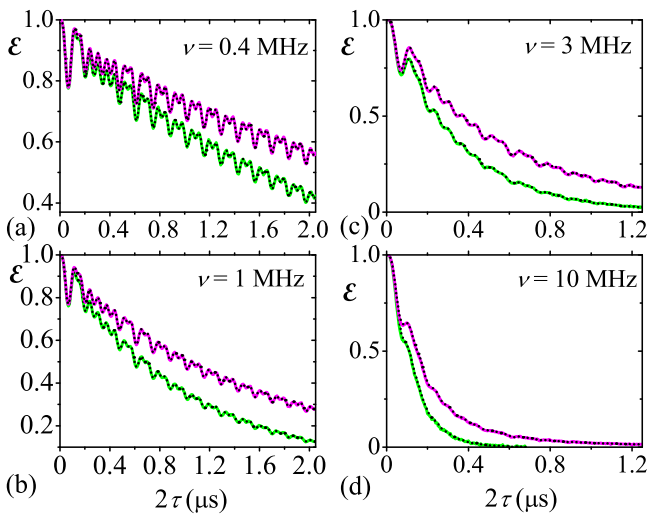}}
\vspace{-0.4cm} \caption{(Color online) The decay of echo
modulation for slow polaron hopping, $\eta \equiv \nu/
\omega_{\text{hf}} \ll 1$. The right-hand sides of Eq.
(\ref{viaprod}) are plotted for $\alpha \equiv T/ T_0 \gg 1$
(green), and $\alpha=2$ (magenta). The corresponding left-hand
sides are plotted with black dotted lines. The hopping attempt
frequencies and corresponding values of $\eta$ are: $\nu=0.4$~MHz,
$\eta =0.0085$ (a), $\nu=1$~MHz, $\eta =0.021$ (b), $\nu=3$~MHz,
$\eta=0.063$ (c), and $\nu=10$~MHz, $\eta =0.21$ (d). The plots
clearly confirm the validity of Eq. (\ref{viaprod}). }
\label{subst}
\end{figure}
%%%%%%%%%%%%%%%%%%

The analysis of $\mathcal{E}(2\tau)$ reveals different types of
$\tau$- dependence in slow ($\eta \ll 1$) and fast ($\eta\gg1$)
hopping regimes. In the slow hopping regime, where this dependence
is more complex, we numerically find that $\mathcal{E}(2\tau)$ is
quite accurately quantified by
\begin{equation}\label{viaprod}
\mathcal{E}(2\tau)= \langle E(2\tau)\rangle
\mathcal{E}_{\text{SC}}(2\tau),
\end{equation}
where $\langle E(2\tau)\rangle $ is established in the previous
Section. To substantiate this relation, in Fig. \ref{subst} we
plot $\mathcal{E}(2\tau)$ numerically calculated for four
different small values of $\eta$, and compare them with the curves
resulting from Eq. (\ref{viaprod}). The plots confirm the validity
of Eq. (\ref{viaprod}) for the hopping attempt frequencies up to
$\eta = 0.21$.

Equation (\ref{viaprod}) suggests that the fine structure of
$\mathcal{E}(2\tau)$ is totally described by $\langle
E(2\tau)\rangle $, whereas its decay is given by
$\mathcal{E}_{\text{SC}}(2\tau)$. Important to us is the question
whether the decay destroys any information on the spectrum of
contact HFI, enclosed in $\langle E(2\tau)\rangle $, i.e., in
$\langle E_c(2\tau)\rangle $. The answer is found from Fig.
\ref{fcfd}(a), indicating that $F_c(\tau)$
%is strongly reduced for $\tau > 0.5\,\mu$s and
almost disappears for $\tau >1\,\mu$s. Thus, one is able to
capture the complete spectrum if $\mathcal{E}(2\tau)$ is
detectable for $\tau\leq 1\,\mu$s.
%while the basic features of spectrum are captured for $\tau\leq 0.5\,\mu$s.
Assuming that $\mathcal{E}(2\tau)\geq 0.05\,\mathcal{E}(0)$ is
%still an observable signal
the restriction for the observation time, we find that for
$\alpha\to\infty$ the contact HFI spectrum is not distorted if
$\nu\leq 1.5$~MHz. At the same time, from Fig. \ref{fcfd}(a) one
can see that $F_c(\tau)$ is essentially non-zero for $\tau \leq
0.5\,\mu$s, meaning that the basic spectral features are
detectable for $\nu\leq 3$~MHz.
% i.e., for $\eta\leq 0.063$.

For $\alpha\to\infty$ and larger $\nu$ the signal decay is faster
and the spectrum distortion is progressively stronger.
Furthermore, in the regime of fast hopping, $\eta\gg 1$, the fine
structure of $\mathcal{E}(2\tau)$ is completely destroyed, even
though the signal decays slower because of the motional narrowing.
Instead of Eq. (\ref{viaprod}), here we get
\begin{equation}\label{triv}
\mathcal{E}(2\tau)= \mathcal{E}_{\text{SC}}(2\tau).
\end{equation}
Therefore, for $\nu > 3$~MHz low-temperature (small-$\alpha$)
measurements can be crucial for the assessment of the primary
ESEEM spectrum.

The experiment Ref. \onlinecite{BoehmePRL12} confirms that the
primary echo signal in MEH-PPV decays exponentially, for at least
$T\geq 10$~K. This experiment does not address the fine structure
of $\mathcal{E}(2\tau)$. However, the results of Ref.
\onlinecite{BoehmePRL12} suggest a uniform polaron hopping;
$\alpha\gg 1$.
%for at least $T>10$~K,
At $T=10$~K the hopping rate is estimated to be
%$W_{\mathbf {r}} \approx 1.64$~MHz,
$\nu \approx 1.64$~MHz, whereas at $T=295$~K it is $\nu \approx
2.87$~MHz. This refers to the slow hopping regime, where the ESEEM
fine structure is shown to be observable.

\subsection{Stimulated ESEEM of hopping polarons}

The stimulated ESEEM of an ensemble of hopping polarons,
$\mathcal{E}(\tau,T)$, is treated in the same way. We introduce
its semiclassical counterpart, $\mathcal{E}_{\text{SC}} (\tau,
T)$, and determine its $T$- dependence. Unlike the above analysis,
however, here we restrict ourselves to the hopping regime, $\eta
<1$, relevant for MEH-PPV.

In the high-temperature limit, $\alpha\to\infty$, we find the
simple exponential decay,
\begin{equation}\label{stEsc}
\mathcal{E}_{\text{SC}}(\tau,T)= \mathcal{E}_{\text{SC}}(2\tau)
\exp(-\nu T).
\end{equation}
For finite $\alpha$ this decay slows down and becomes
non-exponential. Similar to the primary ESEEM, the fine structure
of the stimulated ESEEM is accurately described by the relation,
\begin{equation}\label{steviaprod}
\mathcal{E}(\tau,T)= \langle E(\tau,T)\rangle
\mathcal{E}_{\text{SC}}(\tau,T),
\end{equation}
with $\langle E(\tau,T)\rangle $ characterized in the previous
Section.

The same question as to whether the decay destroys any information
enclosed in $\langle E(\tau, T)\rangle$ on the contact HFI, i.e.,
in $\langle E_c(\tau, T)\rangle$, should be answered in this case.
The question is relevant for stimulated ESEEM measurements aimed
at the detection of the contact HFI, which imply
$\tau=(\pi/\omega_I)n$ with even $n$. The answer is found from
Eqs. (\ref{stEsc}), (\ref{steviaprod}), and the fact that the
amplitude of $\langle E_c(\tau_n,T)\rangle$ is very small for $ T>
0.5\,\mu$s and nearly vanishing for $T> 0.75\,\mu$s (see Appendix
\ref{AppA}). Assuming that the observation time is restricted by
$\mathcal{E}(\tau_n,T)\geq 0.05\,\mathcal{E}(\tau_n,0)$,
%in the limit of large $\alpha$
for $\alpha\gg1$ the complete contact HFI spectrum of the
stimulated ESEEM is detectable for $\nu\leq 4$~MHz, while its
essential spectral features are preserved for $\nu\leq 6$~MHz.
These limits are less restrictive than those on the primary ESEEM
also because the decay of $\mathcal{E}_{\text{SC}}(\tau,T)$ with
$T$ is twice slower than that of $\mathcal{E}_{\text{SC}}(2\tau)$
with $\tau$, cf. Eqs. (\ref{slowsc}) and (\ref{stEsc}).

\section{Concluding remarks}
\label{secDiscuss}

We have studied the ESEEM spectroscopy of polarons in polymer
MEH-PPV. As a reference, we adopted the microscopic picture of the
polaron orbital state and its hyperfine interaction with the
surrounding protons, established in ENDOR and light-induced ESR
experiments. \cite{Kuroda94PRL, Kuroda95, Kuroda2000} Our study
incorporates the random orientations of polymer chains and the
polaron random hopping. The resulting ESEEM spectra have distinct
features from the polaron spin interaction with the distant
protons and from that of the contact protons, which are intrinsic
to the polaron. Utilizing the stimulated ESEEM we formulate a
method which makes the separate observation of the interaction
with the contact protons feasible, by properly choosing the time
parameters.

Electrical or optical detection of any magnetic resonance relies
upon the phenomenon of spin-dependent charge carrier recombination
and transport. Since the work of Kaplan, Solomon, and Mott,
\cite{KSM} the explanation of this phenomenon in terms of weakly
coupled polaron spin pairs is standard.
%An explanation of this phenomenon in terms of weakly coupled
%polaron spin pairs was provided in the early work of Kaplan,
%Solomon, and Mott. \cite{KSM}
Accordingly, pEDMR based ESEEM studies should include
%measurements can be sensitive to the
a weak polaron-polaron spin coupling. The perturbatively
established effect of polaron pair spin coupling on ESEEM
\cite{ZH} consists in the partial shifts of modulation
frequencies, $\delta\omega_{\pm}\approx \pm (J+D)^2/\omega_I$,
where $J$ and $D$ are the strengths of the polaron pair spin
exchange and dipolar couplings, respectively. In the case of
MEH-PPV remote polaron pairs it is reasonable to neglect the spin
exchange.
%because the polaron wavefunctions do not overlap.
As for the dipolar coupling, its contribution can be neglected if
$D^2/\omega_I\ll\omega_I$. In the adopted model of polaron this
condition is met for the polaron separation greater than 2~nm. In
our consideration we neglected the effect of polaron-polaron spin
coupling, assuming such large inter-polaron distances.

In a conventional ESR experiment the echo modulation is subject to
a relaxation decay due to, e. g., electron-nuclear, spin-lattice,
and dipole-dipole interactions. In addition to these, in pulsed
ODMR and EDMR experiments various recombination-dissociation
processes can also contribute in the ESEEM decay. However, the
decay timescales measured so far \cite{Dane08, BoehmePRL12,
Kipp15} point that the polaron hopping yields the fastest channel
of decay. We address the destructive effect of the polaron hopping
and determine the hopping regimes where the ESEEM spectral
features are not distorted. Based on the experiment
Ref.~\onlinecite{BoehmePRL12} we conclude that the polaron hopping
in MEH-PPV is within this regime and our method is viable.

A pulsed EDMR study of the stimulated ESEEM spectrum of polarons
in MEH-PPV was recently reported. \cite{Malissa} Apparently, the
working point in Ref. \onlinecite{Malissa} is such that the
spectral peak from the distant protons is strongly dominant. We
believe that choosing the parameters as proposed above can lead to
the detection of contact hyperfine spectrum, within the same
experimental setup.

Our theory provides the means for further adjustments to the
adopted picture of MEH-PPV polaron
%orbital state, its hyperfine interaction, and transport properties
hyperfine interaction and transport properties, provided
measurements according to the proposed method are available.
Moreover, the theory can be straightforwardly generalized for
organic materials lacking coherent charge transport, other than
MEH-PPV.

\section*{Acknowledgments}

We thank J. Shinar, M. E. Raikh, C. Boehme, H. Malissa, and  M.~E.
Flatt\'{e} for helpful discussions. Work at the Ames Laboratory
was supported by the US Department of Energy, Office of Science,
Basic Energy Sciences, Division of Materials Sciences and
Engineering. The Ames Laboratory is operated for the US Department
of Energy by Iowa State University under Contract No.
DE-AC02-07CH11358.

%This work was supported by the Department of Energy-Basic
%Energy Sciences under Contract No. DE-AC02-07CH11358

\appendix

\section{}

\label{AppA}

In this Appendix we describe the details of the theoretical
framework for the analysis in Section \ref{SecOrDis}.
Particularly, we address the disorder-averaged time-domain
modulation signals $\langle E(2\tau) \rangle$, $\langle E(\tau,T)
\rangle$, and their spectral functions, $\tilde{E}(\omega)$ and $
\tilde{E}(\tau, \omega)$, in line with Ref. \onlinecite{DT}.

In real experiments, as well as during numerical simulations,
time-domain signals are found at discrete values of time.
Typically, one obtains an array of values, $f(t_k)$, for
equidistant time points, $t_k= k\Delta t$, $k=0,1,..,L$. For the
spectral analysis of such a signal it is convenient to introduce
the discrete cosine Fourier transform, $\mathcal{F}_t[f(t)]
(\omega) \equiv \tilde{f}(\omega)$, as
\begin{equation}\label{dcFt}
\tilde{f}(\omega_j) = \sum_{k=0}^L 2f(t_k)\cos(\omega_jt_k)-
f(t_0) + f(t_L),
\end{equation}
where $\omega_j= j\Delta\omega$ with $\Delta\omega = 2\pi/(\Delta
t[L+1])$ and integer $j$, while the last two terms are included to
ensure a zero background. Because of the symmetry,
$\tilde{f}(\omega_j) = \tilde{f}(2\pi/\Delta t-\omega_j)$, it is
appropriate to confine $0\leq j\leq L/2$, restricting the
frequency domain to $0\leq \omega < \pi/\Delta t$. Without going
into the details we assume $\Delta t$ small enough to cover the
necessary frequencies, and $L$ large enough to ensure small
frequency steps. Then one can regard $\tilde{f}$ as a function of
continuous $\omega$. This defines the cosine Fourier transforms we
employ for the spectral analysis of modulation signals:
\begin{equation}\label{prcFt}
\tilde{E}(\omega) = \mathcal{F}_\tau[ \langle E(2\tau)
\rangle],\quad \tilde{E}(\tau, \omega) = \mathcal{F}_T [ \langle
E(\tau,T) \rangle].
\end{equation}

%%%%%%%%%%%%%%%%%%%
\begin{figure}[t]
\vspace{-0.5cm}
\centerline{\includegraphics[width=95mm,angle=0,clip]{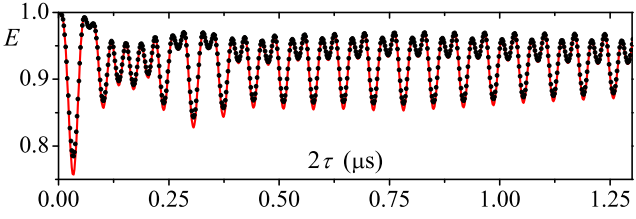}}
\vspace{-0.5cm} \caption{(Color online) Comparison of the polymer
chain orientation disorder averaged exact relation Eq.
(\ref{ndprE}) (black points) and approximation Eq. (\ref{expand})
(red line). } \label{compare}
\end{figure}
%%%%%%%%%%%%%%%%%%

Direct numerical evaluation of modulation depths
%in the adopted model of polaron
from Eq. (\ref{moddepth}) shows that, for all orientations of the
%directions of the applied magnetic field
polymer chains, the maximum modulation depth of the contact
hyperfine protons is
%$k_{c,\text{max}}$
$0.05$ and the maximum depth of the distant protons is
%$k_{d,\text{max}}$
$0.007$ (recall that, for  MEH--PPV, in Eqs. (\ref{ndprE}) and
(\ref{ndstE}) the contact protons are labelled by the subscript,
$1 \leq j\leq N_c$, where $N_c = 22$, and the distant protons are
labelled by $N_c < j\leq N$). This allows to approximate the
factors in Eqs. (\ref{ndprE}) and (\ref{ndstE}) with exponents.
For the primary ESEEM one gets:
\begin{equation}\label{EviaExp}
E(2\tau)= \exp\! \left[ \!-\!\sum_{j=1}^N\! 2k_j\sin^2\! \left(
\frac{\omega_{j +}\tau}2 \right)\! \sin^2\! \left( \frac{\omega_{j
-}\tau}2\right)\! \right]\! .
\end{equation}
To some extent, the argument in Eq. (\ref{EviaExp}) is
characterized by the sum of all depths, $\kappa=\sum_{j=1}^N k_j$.
With the polymer orientation, $\kappa$ varies between 0.03 and
0.242, and averages at about 0.136. The contribution of distant
protons in this sum, $\kappa_d= \sum_{j>N_c} k_j$, is less than
0.06, with the average over the orientation disorder,
$\langle\kappa_d \rangle = 0.047$. Dominant in $\kappa$ is the
contribution of contact hyperfine protons, $\kappa_c=
\sum_{j=1}^{N_c} k_j$, which has a maximum of 0.2 and averages at
about 0.089. However, the contact hyperfine protons have a large
dispersion of modulation frequencies, and even relatively large
fluctuations of $\kappa_c$ do not generate a large argument in Eq.
(\ref{EviaExp}). Therefore it is reasonable to expand the exponent
(\ref{EviaExp}) and write:
\begin{equation}\label{expand}
E(2\tau)\approx 1 -\sum_{j=1}^N 2k_j\sin^2\! \left(
\frac{\omega_{j +}\tau}2 \right)\! \sin^2\! \left( \frac{\omega_{j
-}\tau}2\right).
\end{equation}
This approximation is further reinforced by averaging Eqs.
(\ref{ndprE}) and (\ref{expand}) over orientation disorder
numerically and comparing the results in Fig. \ref{compare}. After
a simple transformation Eq. (\ref{expand}) goes into
Eq.(\ref{aprprE}) of the main text.

The approximation Eq. (\ref{ompmapr}) in the main text for the
distant protons is based on the fact that the polaron spin
coupling to these protons is weak, $A_j, B_j \ll \omega_I$. The
following arguments substantiate the same approximation for the
contact protons. From Eq. (\ref{endorfq}) it is seen that the
approximation error is $\propto B_j^2/(\omega_I \pm A_j/2)$.
Consistent with this, we numerically find the largest error,
$\omega_{j +}-(\omega_I + A_j/2)\approx 0.1$~MHz, occurring for
the largest $B_j$. It results for a C--H proton at vinyl site E,
when the external magnetic field is
%in the plane of polymer chain, directed
parallel to $\hat{\mathbf{x}} + \hat{\mathbf{y}}$ in the principal
axes at E (see Fig. \ref{ppv}). This error is about $1\%$ of the
corresponding frequency values, so the approximation is quite
accurate.

The stimulated ESEEM is analyzed in a similar way. By virtue of
small values of $k_j$, Eq. (\ref{ndstE}) is reduced to the sum Eq.
(\ref{steapr}) in the main text. After averaging over the disorder
in polymer chain orientations and separating the contact and
distant proton contributions $\langle E_\beta (\tau,T) \rangle$,
$\beta=d$, $c$, one gets
\begin{eqnarray}
\label{stimEcd} &&\langle E_\beta (\tau,T) \rangle = -\frac 12
\kappa_\beta + \frac 12 F_\beta(\tau) \cos \omega_I \tau \nonumber \\
&&\hspace{1.5cm} + \frac 12 F_\beta(\tau + T) \cos \omega_I [\tau + T] \\
&&-\frac 14 F_\beta(T) \cos \omega_I [2\tau+T]-\frac 14
F_\beta(2\tau + T) \cos \omega_I T, \nonumber
\end{eqnarray}
%\begin{widetext}
%\begin{equation}
%\label{stimEcd} \langle E_\beta (\tau,T) \rangle = -\frac 12
%\kappa_\beta + \frac 12 F_\beta(\tau) \cos \omega_I \tau + \frac
%12 F_\beta(\tau + T) \cos \omega_I [\tau + T] -\frac 14 F_\beta(T)
%\cos \omega_I [2\tau+T] -\frac 14 F_\beta(2\tau + T) \cos \omega_I T,
%\end{equation}
%\end{widetext}
from which Eq. (\ref{Tpart}) of the main text is written.

%%%%%%%%%%%%%%%%%%%
\begin{figure}[t]
\vspace{-0.3cm}
\centerline{\includegraphics[width=95mm,angle=0,clip]{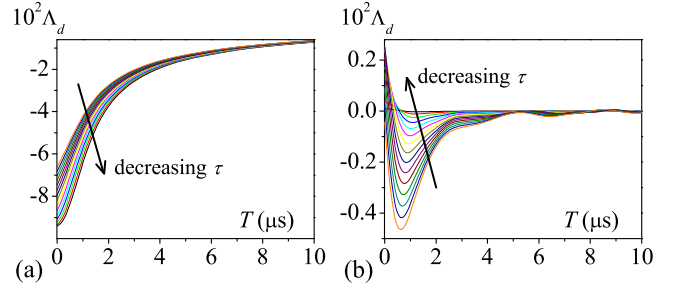}}
\vspace{-0.4cm} \caption{(Color online) Plots of the amplitude,
$\Lambda_d(\tau_n, T)$, versus $T$ at fixed
$\tau_n=(\pi/\omega_I)n$, for (a) odd $n=1,3,..,29$, and (b) even
$n=2,4,..,30$. The plots demonstrate the reduction of
$\Lambda_d(\tau_n, T)$ when going from odd to even $n$. For small
$n$ the decrease of $\Lambda_d(\tau_n, T)$ from odd to even $n$ is
more than two orders of magnitude. For large $n$ it is more that
15 times.} \label{Ld}
\end{figure}
%%%%%%%%%%%%%%%%%%

The $T$- dependence of $\langle E_d (\tau,T)\rangle$ is simple
modulation with the frequency $\omega_I$. To find its $\tau$-
dependence we rewrite the modulation part of Eq. (\ref{stimEcd})
as
\begin{equation}
\label{Lambandphi} \langle E_\beta (\tau,T)\rangle \simeq \frac 12
\Lambda_\beta(\tau,T) \cos\bigl(\omega_I T + \varphi_\beta(\tau,T)
\bigr),
\end{equation}
with $\varphi_\beta = \arg Z_\beta (\text{mod } \pi)$ and
$\Lambda_\beta = Z_\beta e^{-i\varphi_\beta}$, where
\begin{equation}
\label{zmu} Z_\beta = e^{i\omega_I \tau} F_\beta(\tau + T)- \frac
12 e^{2i\omega_I \tau} F_\beta(T) - \frac 12 F_\beta (2\tau + T).
\end{equation}
As defined, $\Lambda_d(\tau,T)$ and $\varphi_d(\tau,T)$ are smooth
functions of $T$, varying insignificantly at times, $T\sim
1/\omega_I$. In contrast, their $\tau$- dependence is abrupt,
because of the presence of exponential factors in Eq. (\ref{zmu}).
The largest and smallest values of $\Lambda_d(\tau,T)$ for a fixed
$T$ can be found in an adiabatic accuracy,
%where $\partial_\tau \Lambda_d(\tau,T)$ is found
by differentiating the fast exponents with respect to $\tau$,
while regarding the $F_d$ factors as constants. It is in fact more
convenient to use the relation, $\Lambda_d^2 =|Z_d|^2$, and
differentiate $|Z_d|^2$. One gets:
\begin{eqnarray}
\label{dzsq} &&\partial_\tau |Z_d|^2 \approx \omega_I\sin \omega_I
\tau\bigl[F_d(\tau + T)F_d(2\tau + T) \\
&& \quad + F_d(T)F_d(\tau + T) - 2\cos \omega_I \tau
F_d(T)F_d(2\tau + T) \bigr]. \nonumber
\end{eqnarray}
This yields minima at $\omega_I\tau_n = \pi n$ for even integer
$n$ and maxima at $\omega_I\tau_n = \pi n$ for odd integer $n$, as
%it was
expected.

%%%%%%%%%%%%%%%%%%%
\begin{figure}[t]
\vspace{-0.3cm}
\centerline{\includegraphics[width=95mm,angle=0,clip]{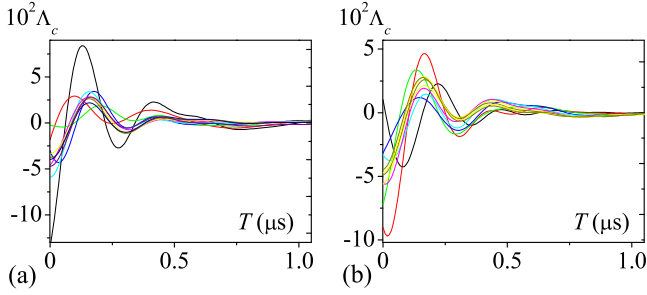}}
\vspace{-0.4cm} \caption{(Color online) The amplitude,
$\Lambda_c(\tau_n, T)$, is plotted versus $T$ at fixed
$\tau_n=(\pi/\omega_I)n$, for (a) odd $n=1,..,19$, and (b) even
$n=2,..,20$. Though the individual curves are not well resolved,
it is seen that there is no notable difference in the orders of
magnitude of $\Lambda_c(\tau_n, T)$ with even and odd $n$. }
\label{Lc}
\end{figure}
%%%%%%%%%%%%%%%%%%

To visualize
%the strength of
the modulation reduction, in Fig. \ref{Ld} we plot
$\Lambda_d(\tau_n,T)$ against $T$ for $n=1,..,30$. We note that
these $\tau_n$ include all possible critical values within the
interval, $0<\tau<1\,\mu s$, which covers the experimentally
available $\tau$- domain, taking into account the decay of the
signal in a real experiment. It is seen that for small $n$ the
reduction is more than two orders of magnitude, and for large $n$
it is more than 15 times.

For the contribution of contact hyperfine protons, $\langle E_c
(\tau,T)\rangle$, the modulation given by Eqs. (\ref{stimEcd}) and
(\ref{Lambandphi}) cannot be interpreted as having a single
frequency, because the function $F_c$, and therefore
$\Lambda_c(\tau,T)$ and $\varphi_c(\tau,T)$, vary abruptly on the
timescale, $T\sim 1/\omega_I$. Still, $\Lambda_c(\tau,T)$ gives
the overall strength of this modulation and it is useful to
inspect this quantity for the above critical values of $\tau$.
Figure~\ref{Lc} plots $\Lambda_c(\tau_n,T)$ versus $T$ for the
first 20 values of $\tau_n$. Overall, the magnitudes of
$\Lambda_c(\tau_n, T)$ in Fig.~\ref{Lc} are close to each other
for even and odd $n$, meaning that there is no reduction of the
corresponding modulation. From Fig.~\ref{Lc} we also infer that
$\Lambda_c(\tau_n,T)$, and therefore $\langle E_c
(\tau_n,T)\rangle$, is very small for $ T> 0.5\,\mu$s, and nearly
vanishes for $T> 0.75\,\mu$s.

%This gives a recipe for a strong reduction of the modulation
%coming from distant protons while preserving the modulation of
%contact protons, if desirable.

\section{}

\label{AppC}

In this Appendix we outline the generalization of
Eqs~(\ref{ndprE}), (\ref{ndstE}) for an ensemble of polarons
hopping over the polymer sites of random orientations.
%It is convenient to
Consider pulse sequences similar to those in Fig. \ref{seqs}(a),
but with unequal delay times; $\pi/2$ - $\tau_1$ - $\pi$ -
$\tau_2$ - $echo$, and $\pi/2$ - $\tau_1$ - $\pi/2$ - $T$ -
$\pi/2$ - $\tau_2$ - $echo$. Using the density matrix formalism,
the modulation functions are
\begin{eqnarray}\label{egen}
&&E(\tau_1,\tau_2)= \mathcal{N}\, \text{Tr}\left[U(\tau_1, \tau_2)
\hat{\rho}(0) U^\dag(\tau_1, \tau_2) S^y\right],  \\
&&E(\tau_1,T,\tau_2)= \mathcal{N}\,\text{Tr}\left[U(\tau_1, T,
\tau_2) \hat{\rho}(0) U^\dag(\tau_1, T, \tau_2)
S^y\right],\nonumber
\end{eqnarray}
where $\hat{\rho} (0)\propto (1/2+S^z)$ is the initial density
operator introduced in Eq. (\ref{eef}) and $\mathcal{N}^{-1} =
\text{Tr}\bigl([3\pi/2] \hat{\rho}(0) [3\pi/2]^\dag S^y\bigr)$ is
the normalization factor. The evolution operators are given by
\begin{eqnarray}\label{evop}
&&U(\tau_1, \tau_2)= e^{-i\tau_2 \tilde{H}}\bigl[\pi\bigr]
e^{-i\tau_1
\tilde{H}}\bigl[ \pi/ 2\bigr], \\
&&U(\tau_1,T,\tau_2)=  e^{-i\tau_2 \tilde{H}}\bigl[ \pi/ 2\bigr]
e^{-iT\tilde{H}}\bigl[\pi/2\bigr] e^{-i\tau_1 \tilde{H}}\bigl[
\pi/ 2\bigr],\nonumber
\end{eqnarray}
where $\tilde{H}$ is Hamiltonian (\ref{locHam}) in the coordinate
system rotating around $\hat{\mathbf{z}}$ with the frequency
$\Omega$. For later reference, we also consider the free induction
decay,
\begin{equation}\label{FID}
F(t)=-\mathcal{N}\, \text{Tr}\left[e^{-it \tilde{H}} \bigl[ \pi/
2\bigr] \hat{\rho}(0)\bigl[ \pi/ 2\bigr]^\dag e^{it
\tilde{H}}S^y\right].
\end{equation}
By taking the traces over the polaron spin space, Eqs.
(\ref{egen}), (\ref{FID}) are reduced to the nuclear spin traces,
involving the nuclear spin Hamiltonians,
\begin{equation}\label{pmHam}
h^{\pm} =  \pm\frac12 \sum_{j=1}^N\bigl( A_j I^z_j + B_j I^x_j
\bigr) - \sum_{j=1}^N \omega_I I^z_j.
\end{equation}
Subsequently, the nuclear spin traces are calculated explicitly.
More specifically, we have:
\begin{widetext}
\begin{eqnarray}\label{Fgen}
&&F(t)= 2^{-N}\text{Tr}_I\!\left[ e^{-i t h^- } e^{i t h^+
}\right] = \prod_{j=1}^Nf_j (t), \\
\label{Epgen} &&E(\tau_1,\tau_2)= 2^{-N}\text{Tr}_I\!\left[e^{-i
\tau_2 h^+ } e^{-i \tau_1 h^- } e^{i \tau_1 h^+ }e^{i \tau_2 h^-
}\right]=\prod_{j=1}^N
\epsilon_j (\tau_1,\tau_2),\\
\label{Esgen} &&E(\tau_1,T, \tau_2)=
2^{-N-1}\text{Tr}_I\!\left[e^{-i (\tau_2+T) h^+ } e^{-i \tau_1 h^-
} e^{i (\tau_1+T) h^+ }e^{i \tau_2 h^- }\right]\! +\bigl(\! +
\leftrightarrow -\! \bigr)^\ast = \frac 12 \! \prod_{j=1}^N \!
\epsilon_j^+ (\tau_1,T,\tau_2) +\bigl(\! + \leftrightarrow -\!
\bigr)^\ast,\quad
\end{eqnarray}
where $(\! + \leftrightarrow -\! )^\ast$ denote the complex
conjugates of previous expressions with the superscripts swapped,
and the functions
\begin{eqnarray}\label{gn}
&&f_j(t) = \cos\frac{\omega_{j+} t}2 \cos \frac{\omega_{j-}t}2 +
\frac{\omega_I^2-A_j^2/4-B_j^2/4} {\omega_{j+} \omega_{j-}}
\sin\frac{\omega_{j+} t}2
\sin \frac{\omega_{j-} t}2,\\
&&\epsilon_j(t_1,t_2) = f_j (t_1-t_2)-2k_j \sin\frac{\omega_{j+}
t_1}2 \sin \frac{\omega_{j-} t_1}2 \sin\frac{\omega_{j+} t_2}2
\sin \frac{\omega_{j-} t_2}2,\nonumber\\
&&\epsilon_j^\pm(t_1,T,t_2)= f_j (t_1-t_2)-2k_j
\sin\frac{\omega_{j\pm} (t_1+T)}2 \sin\frac{\omega_{j\pm}
(t_2+T)}2 \sin \frac{\omega_{j\mp} t_1}2 \sin \frac{\omega_{j\mp}
t_2}2,\nonumber
\end{eqnarray}
\end{widetext}
are introduced, with $\omega_{j\pm}$ and $k_j$ given in Eqs.
(\ref{endorfq}), (\ref{moddepth}).

%%%%%%%%%%%%%%%%%%%
\begin{figure}[b]
%\vspace{-0.3cm}
\centerline{\includegraphics[width=90mm,angle=0]{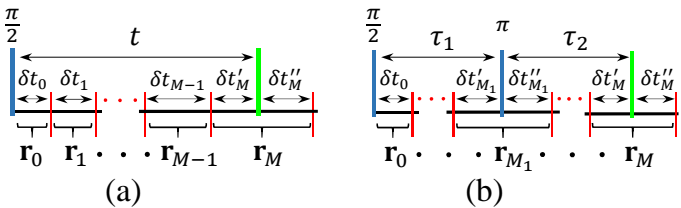}}
\vspace{-0.2cm} \caption{(Color online) Schematic definitions of
polaron random walk trajectories, $\mathbf{R}(t)$, for free
induction decay (a), and primary echo (b). The blue lines denote
the pulses. The green lines show the detection points. The red
bars are polaron random hops. } \label{Rx}
\end{figure}
%%%%%%%%%%%%%%%%%%

To generalize Eqs. (\ref{Fgen}) for hopping polarons, consider a
polaron random walk right after the initial $\pi/2$-pulse (time
$t=0$) from some polymer site, $\mathbf{r}_0$. Its trajectory,
$\mathbf{R}(t)$, specifies
%the moment of time, $t_n$, when the polaron hops from the polymer
%site, $\mathbf{r}_{n-1}= \mathbf{R}(t_n^-)$ to the site, $\mathbf{r}_n
%= \mathbf{R}(t_n^+)$, and the waiting time, $\delta t_n= t_{n+1}-t_n$,
the waiting time, $\delta t_n$, which the polaron spends at
$\mathbf{r}_n$. Other necessary details of $\mathbf{R}(t)$ are
represented in Fig.~\ref{Rx}(a), showing that for time $t$ the
polaron performes $M$ hops, arriving in the site $\mathbf{r}_M$
time $\delta t'_M$ before the detection. The prime indicates that
$\delta t'_M$ is not the total waiting time at $\mathbf{r}_M$. By
this definition,
\begin{equation}\label{tfid}
t= \delta t'_M + \sum_{n=0}^{M-1}\delta t_n, \qquad \delta
t_M=\delta t'_M + \delta t''_M,
\end{equation}
%and $\delta t''_M$ is totally irrelevant to $F_\mathbf{R}(t)$.
The free induction decay of a polaron undergoing such a random
walk is given by
\begin{equation}\label{rwFID}
F_\mathbf{R}(t)=2^{1-N(M+1)}\, \text{Tr}\left[u_\mathbf{R}(t) S^y
u_\mathbf{R}^\dag(t) S^y\right],
\end{equation}
with the time-ordered operator $u_\mathbf{R}(t)$, replacing the
exponential factors in Eq. (\ref{FID}),
\begin{equation}\label{rwevop}
u_\mathbf{R}(t)= e^{-i \delta t'_{M} H_{M}} \prod
 \limits_{n=0}^{{M-1}\atop\longleftarrow} e^{-i \delta t_n H_n}.
\end{equation}
Here the arrow indicates the inverse order of factors in the
products.
%Equation (\ref{rwevop}) implies that for time $t$
The transient Hamiltonians in Eq. (\ref{rwevop}) are
%given by
\begin{equation}\label{tdHam}
H_n = \sum_{j=1}^N S^z\bigl(A_{j,\mathbf{r}_n}
I^z_{j,\mathbf{r}_n} + B_{j,\mathbf{r}_n} I^x_{j,\mathbf{r}_n}
\bigr) - \sum_{l=0}^M\sum_{j=1}^N\omega_I I^z_{j,\mathbf{r}_l},
\end{equation}
where $\mathbf{I}_{j,\mathbf{r}}$ is the spin operator and
$A_{j,\mathbf{r}}$, $B_{j,\mathbf{r}}$ are the hyperfine coupling
constants of the $j$-th proton located at site $\mathbf{r}$, and
the sum over $l$ includes all $M+1$ molecular sites visited for
the random walk $\mathbf{R}(t)$. The time dependence of the spin
Hamiltonian is thus incorporated in the second term of Eq.
(\ref{tdHam}), describing the hyperfine coupling of the polaron
spin with protons near the site, $\mathbf{r}_n= \mathbf{R}(t)$,
occupied by the polaron at time $t$.

The trace over the polaron spin space in Eq.~(\ref{rwFID}) can be
easily taken as the transient Hamiltonians (\ref{tdHam}) conserve
$S^z$. The result is written in terms of the trace over the
nuclear spins:
\begin{equation}\label{FIDviapm}
F_\mathbf{R}(t)= 2^{-N(M+1)} \text{Tr}_I\!\left[u_{\mathbf{R},
-}(t)u_{\mathbf{R},+}^\dag(t) \right].
\end{equation}
where we have introduced
\begin{equation}\label{evoppm}
u_{\mathbf{R},\pm}(t)= e^{-i \delta t'_M h_M^{\pm}} \!\!\prod
 \limits_{n=0}^{{M-1}\atop\longleftarrow} e^{-i \delta t_n h_n
 ^{\pm}}.
\end{equation}
The spin Hamiltonians, $h_n^{\pm}$, are given by
Eq.~(\ref{pmHam}), with the coupling constants and spin operators
of protons at $\mathbf{r}_n$. Note that unlike Eq. (\ref{tdHam}),
the last term in Eq. (\ref{pmHam}) involves nuclear spin operators
only for a single site. This simplification is general for
transport models neglecting the polaron returns to the sites
visited previously, such as the multiple trapping model adopted in
this study. Moreover, neglecting the polaron returns allows to
calculate the trace in Eq. (\ref{FIDviapm}) explicitly. One finds:
\begin{equation}\label{FR}
F_\mathbf{R}(t)= \left( \prod \limits_{n=0}^{M-1}F_n(\delta
t_n)\right)F_M(\delta t'_M),
\end{equation}
where $F_n(t)$ is the free induction decay Eq. (\ref{Fgen})
calculated for the single site, $\mathbf{r}_n$.

Similar expressions can be written for the primary and stimulated
ESE modulation functions, provided the polaron random walk
trajectory is specified relative to the pulse sequence. Namely,
for the primary sequence let $\mathbf{R}(\tau_1+\tau_2) =
\mathbf{r}_M$, and the instantaneous $\pi$-pulse is applied
$\delta t'_{M_1}$ time after the polaron arrives in the site
$\mathbf{r}_{M_1}$, and $\delta t''_{M_1}$ time before it makes
the next hop, see Fig.~\ref{Rx}(b). The primary ESE modulation
from a spin with this trajectory is found to be
\begin{eqnarray}\label{EpR}
E_\mathbf{R}(\tau_1,\tau_2) = \left( \prod
\limits_{n=0}^{M_1-1}F_n(\delta t_n)\right) \! E_{M_1}
(\delta t'_{M_1},\delta t''_{M_1} )\quad && \\
\times   \left(\prod \limits_{n=M_1+1}^{M-1} F_n(\delta t_n)
\right) \! F_M(\delta t'_M) ,&& \nonumber
\end{eqnarray}
where $E_n(t_1,t_2)$ is the modulation function (\ref{Epgen}), for
$\mathbf{r}_n$.

\begin{figure}[t]
\vspace{-0.3cm}
\centerline{\includegraphics[width=80mm,angle=0]{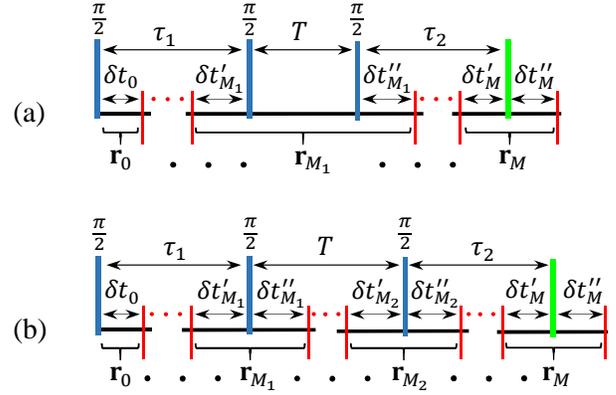}}
\vspace{-0.2cm} \caption{(Color online) Illustration of notations
for polaron random walk trajectories during the stimulated pulse
sequence. The blue bars symbolize the pulses. The green bars show
the detection. The red lines are polaron random hops. (a)
Trajectories of type $\mathbf{R}_0$,  Eq. (\ref{EsR0}); no polaron
hop occurs in the time interval $T$. (b)  Trajectories of type
$\mathbf{R}_1$,  Eq.~(\ref{EsR0}); at least one hop occurs in the
time interval $T$. } \label{Rste}
\end{figure}

The stimulated ESE modulation critically depends on whether a
random walk involves a hop in the interval $T$ or not. We separate
these cases in Fig.~\ref{Rste}(a) and (b). The trajectories with
no hops during the interval $T$, Fig.~\ref{Rste}(a), are denoted
by $\mathbf{R}_0$, while those incorporating hops in $T$,
Fig.~\ref{Rste}(b), by $\mathbf{R}_1$. With the further details of
trajectories specified in Fig.~\ref{Rste}, one gets:
\begin{eqnarray}\label{EsR0}
E_{\mathbf{R}_0}(\tau_1,T,\tau_2) \!\!\!\! &&= \left( \prod
\limits_{n=0}^{M_1-1}F_n(\delta t_n)\right)\!
E_{M_1}(\delta t'_{M_1},T,\delta t''_{M_1} )  \nonumber\\
&&\times  \left(\prod \limits_{n=M_1+1}^{M-1} F_n(\delta t_n)
\right)\! F_M(\delta t'_M),
\end{eqnarray}
where $E_n(t_1,T,t_2)$ is is given by Eq. (\ref{Esgen}) at
$\mathbf{r}_n$, and
\begin{eqnarray}\label{EsR1}
E_{\mathbf{R}_1}(\tau_1,T,\tau_2)= \left( \prod
\limits_{n=0}^{M_1-1}F_n(\delta t_n)\right)F_{M_1}(\delta
t'_{M_1})&& \nonumber \\
\times F_{M_2}(\delta t''_{M_2})\left(\prod \limits_{n=M_2+1}^{M
-1} F_n(\delta t_n) \right) \! F_M(\delta t'_M) . &&
\end{eqnarray}

%%%%%%%%%%%%%%%%%%%
%%%%%%%%%%%%%%%%%%

Finally, the free induction decay and ESE modulations of the
ensemble of randomly walking polarons is found from Eqs.
(\ref{FR}) -- (\ref{EsR1}), via averaging over the random-walk
trajectories:
\begin{eqnarray}\label{rwem1}
&&\mathcal{F}(t) = \left\langle F_\mathbf{R}(t)
\right\rangle_{\{\mathbf{R}\}},\\
\label{rwem2}&&\mathcal{E}(\tau_1, \tau_2) = \left\langle
E_\mathbf{R}(\tau_1, \tau_2) \right\rangle_{\{\mathbf{R}\}},\\
\label{rwem3}&&\mathcal{E}(\tau_1,T, \tau_2) = \left\langle
E_\mathbf{R}(\tau_1, T, \tau_2) \right\rangle_{\{\mathbf{R}\}}.
\end{eqnarray}
The averages are evaluated numerically, by a Monte Carlo sampling
of random walk trajectories, including the random on-site trapping
energies defining the waiting time statistics via Eq. (\ref{MTM}).
In our simulations we also incorporate the random orientations of
polymer chains.

\section{}

\label{AppB}

In this Appendix we investigate $\mathcal{F}(t)$,
$\mathcal{E}(\tau, T)$, and $\mathcal{E}(\tau_1,T, \tau_2)$
analytically, within the multiple trapping model at
$\alpha\to\infty$. This implies uniform hopping rates,
$W_{\mathbf{r}}=\nu$, entailing the Poissonian waiting time
distribution, $P(\delta t)=\nu\exp(-\nu \delta t)$. In this limit
the free induction decay satisfies a Dyson-type integral equation,
\cite{Kubo79, Allodi14}
\begin{equation}\label{gtinteq}
\mathcal{F}(t) = g(t)e^{-\nu t} + \nu \int_0^t\!\! dt' e^{-\nu t'}
g(t') \mathcal{F}(t-t'),
\end{equation}
where the on-site relaxation function,
\begin{equation}\label{g0t}
g(t)= \bigl\langle F(t) \bigr \rangle,
\end{equation}
is introduced. Here, $F(t)$ is given by Eq. (\ref{Fgen}), and the
brackets mean the average over random orientations of molecular
sites. In Eq. (\ref{gtinteq}) the first term is the relaxation if
for time $t$ the polarons do not hop, which occurs with the
probability $e^{-\nu t}$, and the integral accounts for the
relaxation with the first hop happening at time $t'< t$.

The formal solution of Eq. (\ref{gtinteq}) is given in terms of
the Laplace transform:
\begin{equation}\label{LtG}
\tilde{\mathcal{F}}(s)=\frac{\tilde{g}(s+\nu)}{1 - \nu \tilde{g}
(s+\nu)},
\end{equation}
where $\tilde{f}(s)=\int_0^\infty\exp(-st) f(t)dt$ denotes the
Laplace transform of $f(t)$. However, from this equation
$\mathcal{F}(t)$ can be found only numerically, as the inverse
Laplace transform of Eq. (\ref{LtG}) is not accessible
analytically.

\subsection*{Semiclassical description}

A semiclassical approximation for $\mathcal{F}$ and $\mathcal{E}$
follows upon replacing the Hamiltonian in Eqs. (\ref{evop}),
(\ref{FID}) by its semiclassical counterpart, Eq. (\ref{QC}). The
resulting on-site free induction decay has the simple form,
\begin{equation}\label{scg}
g_0(t) = \langle\cos(\omega_z t)\rangle_{\omega_z}=
\exp\bigl(-\omega_{\text{hf}}^2t^2/2\bigr).
\end{equation}
Still, the solution for the semiclassical free induction decay,
$\mathcal{F}_0(t)$, using the inverse Laplace transform
(\ref{LtG}), can be found only numerically. \cite{Allodi14}

In what follows we give a perturbative treatment for the
semiclassical echo modulation functions,
$\mathcal{E}_{\text{SC}}(2\tau)= \mathcal{E}(\tau, \tau)$ and
$\mathcal{E}_{\text{SC}}(\tau,T)= \mathcal{E}(\tau, T, \tau)$,
from which Eqs. (\ref{slowsc}) and (\ref{stEsc}) of the main text
result. In the semiclassical approximation and within the multiple
trapping model at $\alpha\to\infty$, Eqs. (\ref{rwem1}) --
(\ref{rwem3}) can be relates as
\begin{eqnarray}\label{scEpr}
&&\mathcal{E}_{\text{SC}}(2\tau) = e^{-2\nu\tau}\left[1+ 2\nu
\int_0^\tau e^{2\nu t}\mathcal{F}_0^{\,2}(t) dt \right],\\
\label{scEstm} &&\mathcal{E}_{\text{SC}}(\tau,T) = e^{-\nu
T}\mathcal{E}_{\text{SC}}(2\tau) + \mathcal{F}_0^{\,2}(\tau)
\left(1-e^{-\nu T} \right),\quad\qquad
\end{eqnarray}
detailed derivation of which will be given elsewhere. \cite{MDunp}
Thus, $\mathcal{E}_{\text{SC}}(2\tau)$ and
$\mathcal{E}_{\text{SC}} (\tau,T)$ are determined by
$\mathcal{F}_0(t)$. Note that the first term in Eq. (\ref{scEstm})
is the contribution of type $\mathbf{R}_0$ trajectories, Fig.
\ref{Rste}(a), while the last term is that of the type
$\mathbf{R}_1$ trajectories, Fig. \ref{Rste}(b).

In the regime of slow hopping, $\eta\equiv\nu/\omega_{\text{hf}}
\ll 1$, a reasonably good approximation can be made for
$\mathcal{F}_0(t)$ from Eq. (\ref{gtinteq}) iteratively. To the
linear order in $\eta$ one gets:
\begin{equation}\label{iter}
\mathcal{F}_0(t) = e^{-\nu t} \left [g_0(t) + \nu \int_0^t\!\! dt'
g_0(t') g_0(t-t')\right].
\end{equation}
Using this in Eq. (\ref{scEpr}) leads to Eq. (\ref{slowsc}) in the
main text. Equation (\ref{iter}) also shows that the decay of
$\mathcal{F}_0(t)$ is nearly Gaussian and fast, so that for $\tau
>1/ \omega_{\text{hf}}$ the last term in Eq. (\ref{scEstm}) can be
neglected, and Eq. (\ref{stEsc}) in the main text can be written.

In the fast hopping regime, $\eta\gg 1$, the Laplace transform
appears to be useful. One has:
\begin{equation}\label{tdG}
\mathcal{F}_0(t) = \frac 1 {2\pi i}\int_{-i\infty}^{i\infty}
\!\!\! ds\, e^{st} \tilde{\mathcal{F}}_0(s),
\end{equation}
with $\tilde{\mathcal{F}}_0(s)$ given by Eq. (\ref{LtG}) and the
Laplace transform,
\begin{equation}\label{Lgviaerf}
\tilde{g}_0(s) = \sqrt{\pi/ 2}\,\omega_{\text{hf}}^{-1}\exp \bigl(
s^2/2\omega_{\text{hf}}^2\bigr)\text{erfc}\bigl(s/\sqrt{2}\omega_{\text{hf}}
\bigr),
\end{equation}
where $\text{erfc}(x)$ is the complementary error function.
$\tilde{\mathcal{F}}_0(s)$ is holomorphic on the complex
half-plane, $\text{Re}(s)<0$, excluding the simple poles
determined by the denominator of Eq. (\ref{LtG}). A thorough
analysis of the inverse Laplace transform (\ref{tdG}) shows that
$\tilde{\mathcal{F}}_0(s)$ has one real negative pole, $s_0$, and
infinitely many complex poles. \cite{MDunp} Also, for $\eta\gg 1$
the contribution of $s_0$ dominates in the integral (\ref{tdG}),
giving $\mathcal{F}_0(t) = -( \omega_{\text{hf}}^2/\nu
s_0)\exp(s_0t)$. From the large-argument asymptote of Eq.
(\ref{Lgviaerf}) one finds $s_0=-\omega_{\text{hf}}^2 /\nu$,
leading to the well-known result in the motional narrowing regime,
$\mathcal{F}_0(t) = \exp(-\omega_{\text{hf}}^2\, t/\nu)$. With
this $\mathcal{F}_0(t)$, the integral term in Eq. (\ref{scEpr}) is
dominant, yielding $\mathcal{E}_{\text{SC}}(2\tau) =
\exp(-2\,\omega_{\text{hf}}^2\, \tau/\nu)$.


\begin{references}
\bibitem{ftnt} Throughout the paper, by the Fourier transform of a function of
two variables, $(\tau, T)$, we mean the transformation over the
variable $T$, whereas $\tau$ is kept as a parameter. For the
precise definition of these Fourier transforms see Appendix
\ref{AppA}.
\bibitem{ftnt2} The largest value of $|A_j|$ within the HFI under
consideration is about $2\pi\times 9.4$~MHz, whereas the X -- band
proton Larmor frequency is about $\omega_I = 2\pi\times 14.7$~MHz.
\end{references}
\end{document}